\documentclass[aps,prd,preprint,superscriptaddress,tightenlines,nofootinbib]{revtex4}

\usepackage{graphicx}
\usepackage{dcolumn}
\usepackage{bm}

\newcommand{\FigsDir}{.}
\newcommand{\FigsAch}{.}
\newcommand{\FigsSch}{.}
\newcommand{\PM}{$\pm$}

\begin{document}


\preprint{CLNS 07/1993}       
\preprint{CLEO 07-3}          

\title{Dalitz plot analysis of the $D^+\to\pi^-\pi^+\pi^+$ decay}


\author{G.~Bonvicini}
\author{D.~Cinabro}
\author{M.~Dubrovin}
\author{A.~Lincoln}
\affiliation{Wayne State University, Detroit, Michigan 48202}
\author{D.~M.~Asner}
\author{K.~W.~Edwards}
\author{P.~Naik}
\affiliation{Carleton University, Ottawa, Ontario, Canada K1S 5B6}
\author{R.~A.~Briere}
\author{T.~Ferguson}
\author{G.~Tatishvili}
\author{H.~Vogel}
\author{M.~E.~Watkins}
\affiliation{Carnegie Mellon University, Pittsburgh, Pennsylvania 15213}
% ...
\collaboration{CLEO Collaboration} 
\noaffiliation


\date{July 2, 2007}

\begin{abstract}
Using 281~pb$^{-1}$ of data recorded by the CLEO-c detector in
$e^+e^-$ collisions at the $\psi(3770)$, corresponding to 0.78 million 
$D^+D^-$ pairs, we investigate the substructure of the decay $D^+ \to \pi^-\pi^+\pi^+$
using the Dalitz plot technique.
We find that our data are consistent with the following intermediate states:
$\rho(770)\pi^+$,
$f_2(1270)\pi^+$,
$f_0(1370)\pi^+$,
$f_0(1500)\pi^+$,
$f_0(980)\pi^+$, 
and
$\sigma\pi^+$.  
We confirm large S wave contributions at low $\pi\pi$ mass.
We set upper limits on contributions of other possible
intermediate states.  We consider three models of the $\pi\pi$ S wave
and find that all of them adequately describe our data.
\end{abstract}

\maketitle

\section{Introduction}	

The study of charmed meson hadronic decays 
illuminates light meson spectroscopy.  
Many of these decays proceed via quasi two-body modes 
and are subsequently observed as three or more stable particles. 
In this work our goal is
to describe the two-body resonances that contribute
to the observed three-body $D^+ \to \pi^-\pi^+\pi^+$ decay.  
Study of a given state can shed light on different production mechanisms.

We present here a study of charged $D$ decay to three charged pions 
carried out with the CLEO detector. This mode has been studied
previously by E687~\cite{E687_Dp-pipipi}, E691~\cite{E691_Dp-pipipi},
E791~\cite{E791_Dp-pipipi}, and
FOCUS~\cite{FOCUS_Dp-pipipi}.  The analyses from E791 and FOCUS
have roughly the same data size as the one described here, 
while the E687 and E691 analyses used about an order
of magnitude smaller samples and are not discussed further.

E791 uses the isobar technique, where
each resonant contribution to the Dalitz plot~\cite{dalitz}
is modeled as a Breit-Wigner amplitude with a complex phase.
This works well for narrow, well separated resonances, but
when the resonances are wide and start to overlap,
solutions become ambiguous, and unitarity is violated.
In contrast, FOCUS uses the K-matrix approach,
which gives a description of S wave $\pi\pi$
resonances treating the $\sigma$ (also known as $f_0(600)$) and $f_0(980)$ contributions
in a unified way.  
While this approach is a step forward, some authors 
\cite{Oller_2005},
\cite{Bugg_2005_Kpi}
have claimed that the exact formalism used by FOCUS violates chiral 
constraints, and might therefore lead to unphysical
behavior at low $\pi\pi$ mass, where the S wave is most prominent.
Despite the difference in approach
the two techniques give a good description of
the observed Dalitz plots and agree about the
overall contributions of the resonances,
as is shown in Table~\ref{tab:compare1}.
Both experiments see that about half of the fit fraction
for this decay is explained by a low $\pi^+\pi^-$ mass
S wave. 
We have in hand a comparable sample of $D^+ \to \pi^-\pi^+\pi^+$
decays (inclusion of
the charge-conjugate mode is always implicit); 
we can thus check this somewhat surprising result
in a significantly different environment.

E791 and FOCUS are fixed target experiments where
$D$ mesons are produced within a momentum range of 10-100~GeV/$c$.
In our experiment $D^+$ mesons are produced in the process
$e^+e^- \to \psi(3770) \to D^+D^-$,
close to the threshold, and are thus almost at rest.
This difference of production environments is important
for observation of events from the decay $D^+ \to K^0_S\pi^+$,
which has a large rate and contributes to the same final state.
These events are easily removed in the fixed target experiments
by requiring all three charged pions to be consistent with
a common vertex, and its residual contribution was
estimated to be small.  We are forced to take a different
approach as the lower momentum $K^0_S$
does not produce clearly detached vertexes when $K^0_S \to \pi^+\pi^-$.
Nevertheless we are able to clearly isolate
the $K^0_S \pi^+$ channel, using the $\pi^+\pi^-$ invariant mass.

\begin{table}
\caption{A comparison of the observed fit fractions in \% from previous
         studies of $D^+ \to \pi^-\pi^+\pi^+$. The sum of all
         fit fractions is not necessarily equal to 100\% due to the ignored 
         interference terms.        
         The ``S wave $\pi^+$'' entry
         for E791 is the sum of the three entries above it.}
\begin{tabular}{c|c|c}
\hline
\hline
Mode                & E791~\cite{E791_Dp-pipipi}& FOCUS~\cite{FOCUS_Dp-pipipi} \\ 
\hline
$\sigma \pi^+$      & $46.3 \pm 9.2$ & \\
$f_0(980) \pi^+$    & $ 6.2 \pm 1.4$ & \\
$f_0(1370) \pi^+$   & $ 2.3 \pm 1.7$ & \\ 
\hline
S wave $\pi^+$      & $54.8 \pm 9.5$ & $56.0 \pm 3.9$ \\
$\rho^0(770) \pi^+$ & $33.6 \pm 3.9$ & $30.8 \pm 3.9$ \\
$f_2(1270) \pi^+$   & $19.4 \pm 2.5$ & $11.7 \pm 1.9$ \\
\hline
\hline
\end{tabular}
\label{tab:compare1}
\end{table}

Our analysis compares several different models for this decay, 
attempting to find the best description.
One is an isobar model where we have included the best description
of the $\sigma$ from Ref.~\cite{Oller_2005} and the Flatt\'e
parameterization for the threshold effects on the
$f_0(980)$~\cite{Flatte}.  We use two other S wave models, both of which
satisfy chiral constraints and respect unitarity.
A model by Schechter and his collaborators (Schechter model)~\cite{Schechter_2005}
is based on the linear sigma model of the chiral symmetric Lagrangian.  
It includes only
the lowest lying $\pi\pi$ S wave resonances, the $\sigma$ and the $f_0(980)$.
A model by Achasov and his collaborators (Achasov model)~\cite{Achasov_D3pi}
is field-theory based and has been
developed to describe scattering experiments.
We compare the results of these three models of the 
resonance contributions to the Dalitz plot
to see if one description is superior to the others
and to understand differences among the models.

In Section~\ref{sec:detector} we briefly describe the 
CLEO-c experiment and the basic
algorithms of event reconstruction.
In Section~\ref{sec:event_selection} we describe
the event selection for the Dalitz 
plot analysis.
The formalism of fitting the observed Dalitz plot, and
systematic cross-checks are given in
Section~\ref{sec:dalitz_plot_analysis}.
Appendix~\ref{sec:appendix} describes in 
detail the two $\pi^+\pi^-$ S wave models
that we use, some of which are
extensions of published theoretical work.
We summarize our results in Section~\ref{sec:summary}.


\section{Detector and experimental technique}
\label{sec:detector}
CLEO-c is a general purpose detector which includes a tracking system
for measuring momenta and specific ionization of charged particles,
a Ring Imaging Cherenkov detector to aid particle identification,
and a CsI calorimeter for detection of electromagnetic showers.
These components are immersed in a magnetic field of 1~T,
provided by a superconducting solenoid, and surrounded by a muon detector.
The CLEO-c detector is described in detail elsewhere~\cite{CLEO-c}.

This analysis utilizes 281~pb$^{-1}$ of data collected on the $\psi(3770)$
resonance at $\sqrt{s}\simeq$3773~MeV at the Cornell Electron Storage Ring,
corresponding to production of about $0.78\times 10^6$ $D^+D^-$ pairs. 
We reconstruct the $D^+ \to \pi^-\pi^+\pi^+$ decay 
using three tracks measured in the tracking system. 
Charged tracks satisfy standard goodness of fit quality requirements~\cite{HadronicBF}.
Pion candidates are required to have specific ionization,
$dE/dx$, in the main drift chamber within four standard
deviations of the expected value for
a pion at the measured momentum.
Tracks coming from the origin must have an impact parameter
with respect to the beam spot 
(in the plane transverse to the beam direction) 
of less than 5~mm.
We do not reconstruct the $K^0_S \to \pi^+\pi^-$ vertex, but the requirement on
pion track impact parameter
removes $\sim$60\% of events with $K^0_S \to \pi^+\pi^-$ decays.
The remaining events from $D^+ \to K^0_S \pi^+$ represent about 
one third of those selected for the Dalitz plot. 

\begin{figure}
  \begin{minipage}[t]{3.0in}
    \includegraphics*[width=3.0in]{\FigsDir/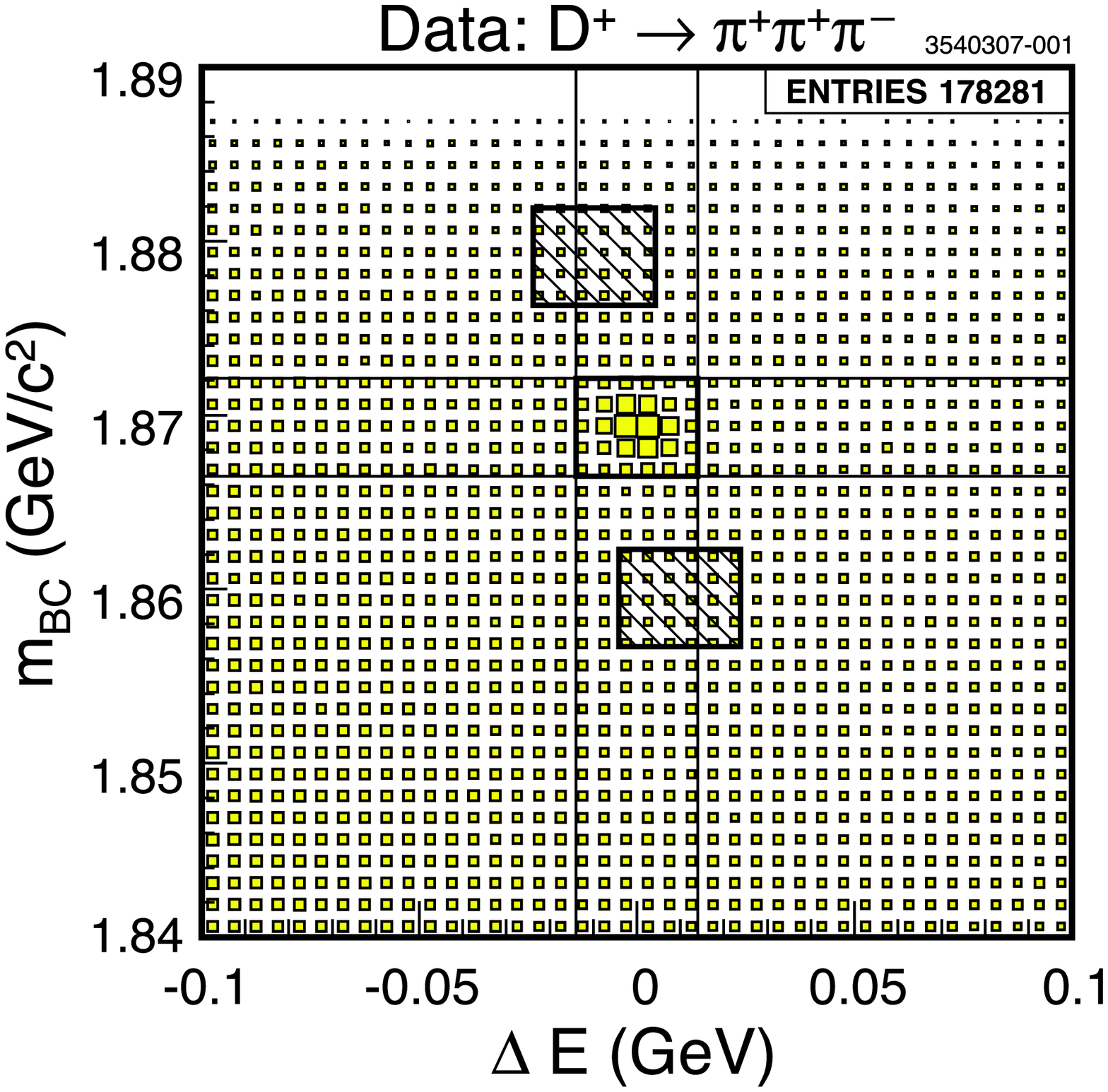}
    \caption{\label{figs:fig_mode_209_dalitz_p1_data} 
             The $m_{\rm{BC}}$ {\sl vs.} $\Delta E$ distribution of events passing
             all selection requirements described in the text. The center box shows
             the signal region for the Dalitz plot analysis.  
             The two hatched boxes show the sidebands. 
	     The vertical and horizontal lines restrict the regions
             of events plotted in 
             Figs.~\ref{figs:fig_mode_209_dalitz_p43_data} and
                  \ref{figs:fig_mode_209_dalitz_p2_data}, respectively.
            }
  \end{minipage}
  \hfill
  \begin{minipage}[t]{3.0in}
    \includegraphics*[width=3.0in]{\FigsDir/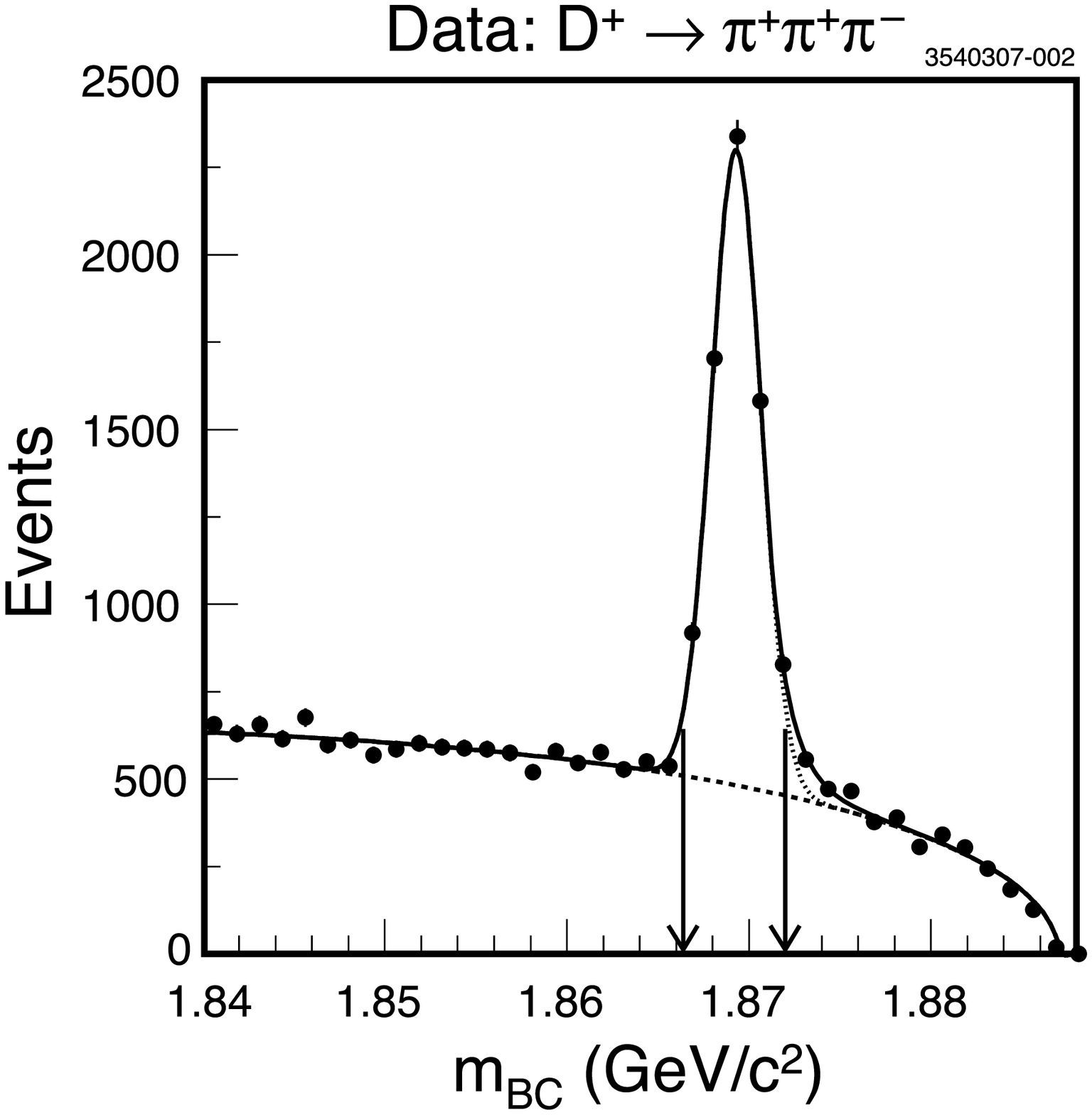}
    \caption{\label{figs:fig_mode_209_dalitz_p43_data}
             The $m_{\rm{BC}}$ distribution of events
             from the $|\Delta E| < 2 \sigma(\Delta E)$ range. 
             Dashed curve shows a contribution from the background,
             dotted curve is a Gaussian part of the Crystal Ball function for the 
             signal shape, and
             solid curve is total, signal plus background.
             Events between arrows are selected for the Dalitz plot analysis.
            }
  \end{minipage}
\end{figure}
\begin{figure}
  \begin{minipage}[t]{3.0in}
    \includegraphics*[width=3.0in]{\FigsDir/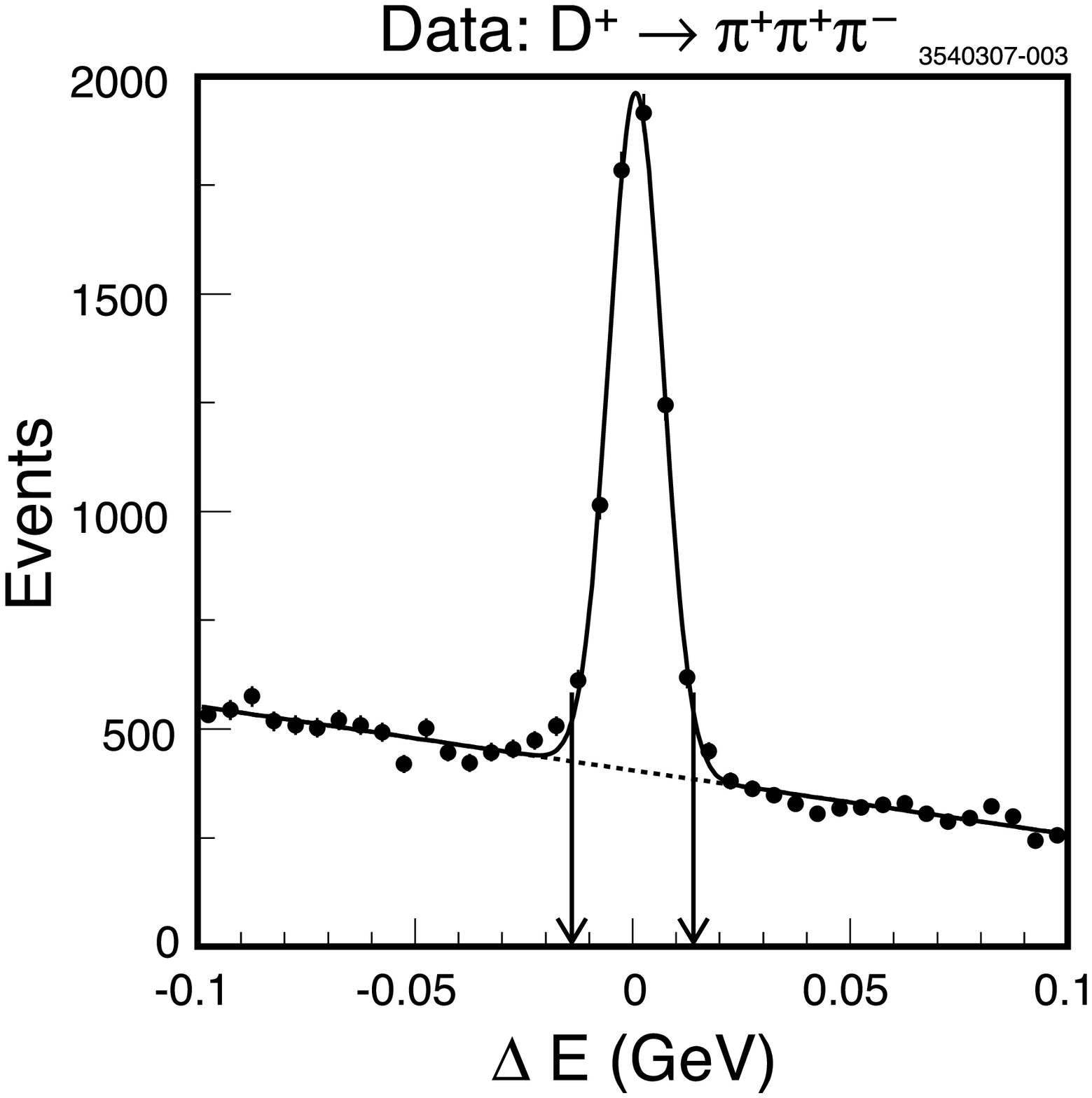}
    \caption{\label{figs:fig_mode_209_dalitz_p2_data}
             The $\Delta E$ distribution of events
             from the $|m_{\rm{BC}} - m_D| < 2 \sigma(m_{\rm{BC}})$ range. 
             Events between the arrows are selected for the Dalitz plot.
            }
  \end{minipage}
  \hfill
  \begin{minipage}[t]{3.0in}
    \includegraphics*[width=3.0in]{\FigsDir/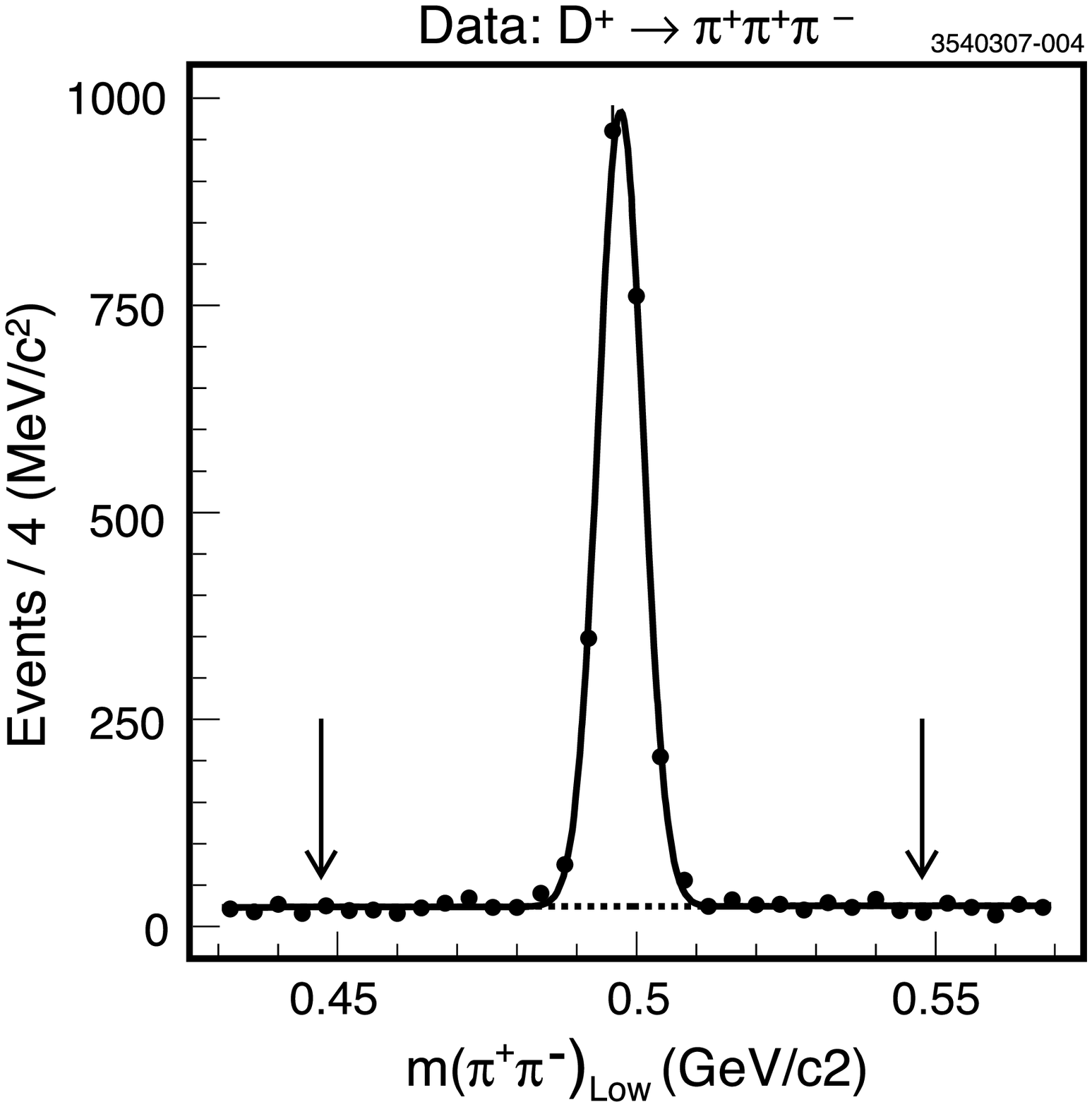}
    \caption{\label{figs:fig_mode_209_dalitz_p12_data}
             The $m(\pi^+\pi^-)_{\rm{Low}}$ distribution of events pre-selected
             for the Dalitz plot. 
             A clear signal for the  $K^0_S \to \pi^+\pi^-$ decay is observed.
             Events in the range between the arrows, 
             $0.2 < m^2(\pi^+\pi^-)_{\rm{Low}} < 0.3$~(GeV/$c^2$)$^2$, 
             are discarded from the Dalitz plot analysis.
             }
  \end{minipage}
\end{figure}


\begin{figure}
  \begin{minipage}[t]{3.0in}
    \includegraphics*[width=3.0in]{\FigsDir/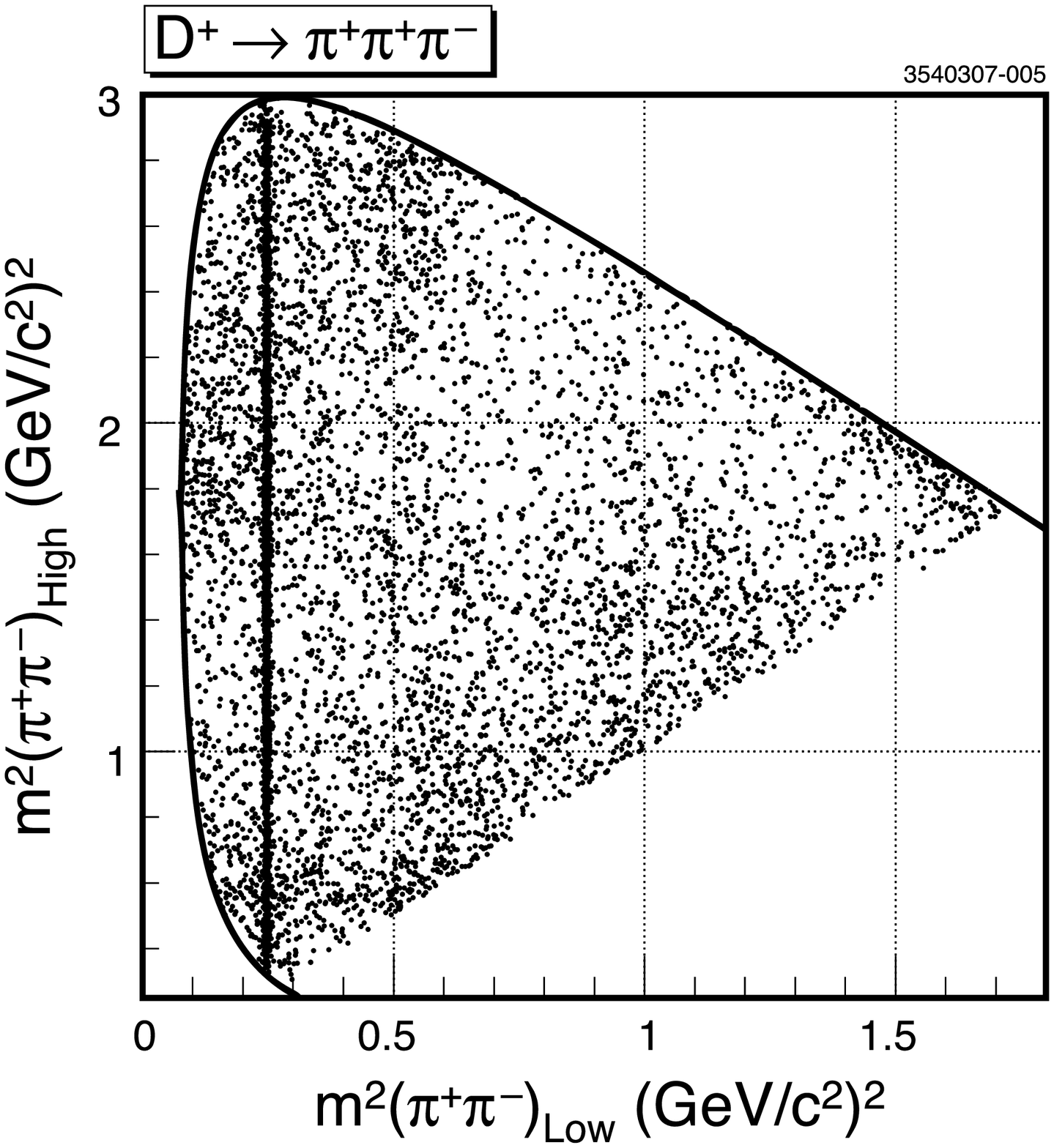}
    \caption{\label{figs:pipipi_sig_DP_dots}
             The Dalitz plot for $D^+ \to \pi^-\pi^+\pi^+$ candidates.
            }
  \end{minipage}
  \hfill
  \begin{minipage}[t]{3.0in}
    \includegraphics*[width=3.0in]{\FigsDir/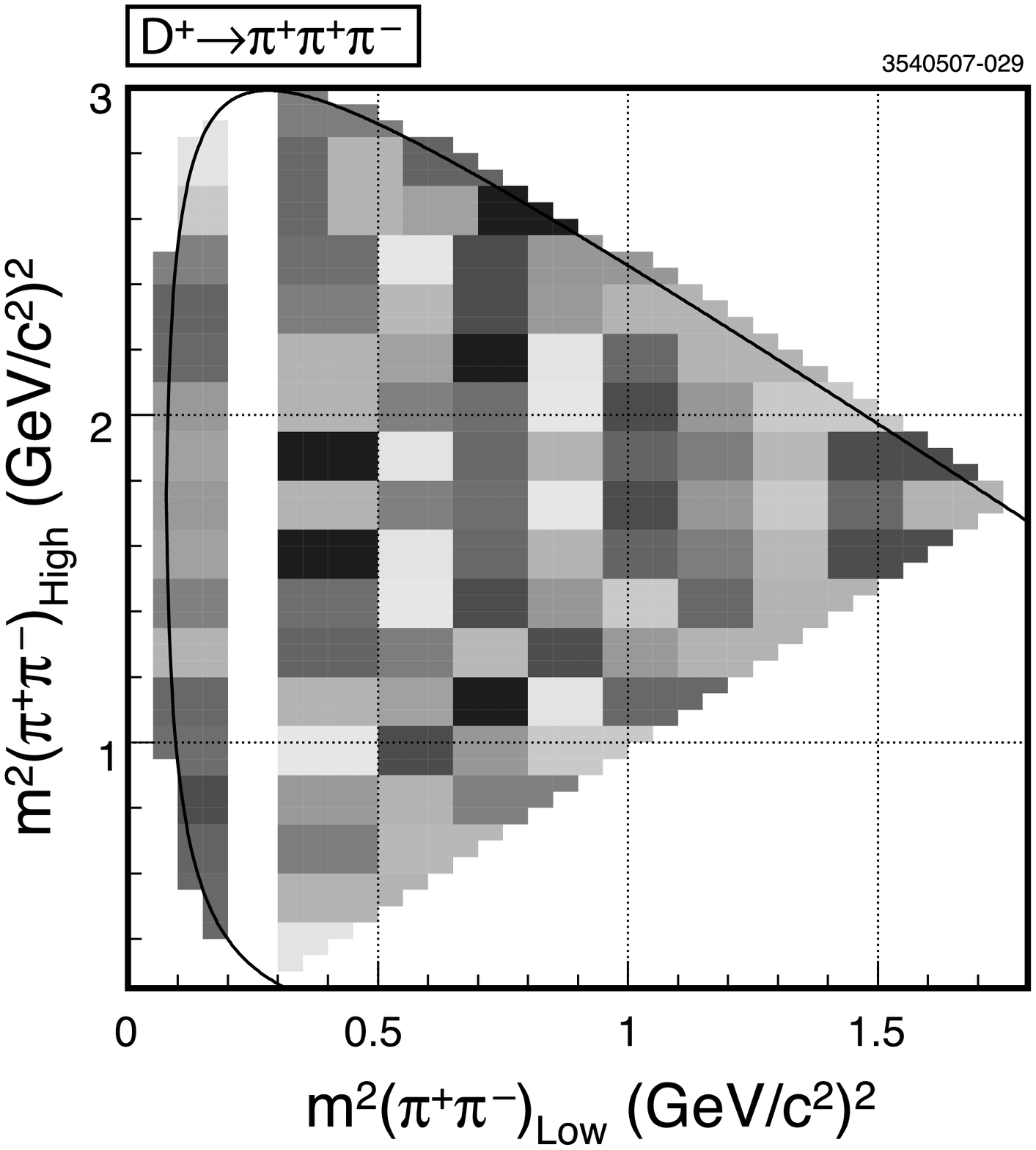}
    \caption{\label{figs:adaptive_binning_scheme}
             The adaptive binning scheme.
            }
  \end{minipage}
\end{figure}

\section{Event selection}
\label{sec:event_selection}

Selection of events from the $D^+ \to \pi^-\pi^+\pi^+$ decay
is done with two signal variables:
\begin{equation}
      \Delta E = E_D - E_{\mathrm{beam}},
\end{equation}
\begin{equation}
      m_{\rm{BC}} = \sqrt{E^2_{\mathrm{beam}}-p^2_D},
\end{equation}
where $E_{\mathrm{beam}}$ is a beam energy, and
$E_D$ and $p_D$ are the energy and momentum of
the reconstructed $D$ meson candidate, respectively.
The beam crossing angle of $\sim$4~mrad 
is used to calculate the $D$ meson candidate energy and momentum
in the $\psi(3770)$ center of mass system.
We require
$|\Delta E| < 2 \sigma(\Delta E)$,
$|m_{\rm{BC}} - m_D| < 2 \sigma(m_{\rm{BC}})$, where resolutions
$\sigma(\Delta E) = 5.5\pm0.4$~MeV and
$\sigma(m_{\rm{BC}}) = 1.38\pm0.03$~MeV/$c^2$ represent the widths of 
the signal peak in the 2D-distribution shown in
Fig.~\ref{figs:fig_mode_209_dalitz_p1_data}, and the projections,
Fig.~\ref{figs:fig_mode_209_dalitz_p43_data} and 
Fig.~\ref{figs:fig_mode_209_dalitz_p2_data}.
To determine the efficiency we use a GEANT-based Monte Carlo
simulation
where one of the charged $D$ meson decays in a signal mode
uniformly in phase space, while the other decays to
all known modes with relevant branching fractions.
Simulated events are
required to pass the same selection requirements as data.
The shape of the background contribution in
the Dalitz analysis is estimated
using events from the two hatched side-band boxes 
shown in Fig.~\ref{figs:fig_mode_209_dalitz_p1_data}. 
The sideband boxes are shifted in $\Delta E$ 
to select the background events whose $\pi^-\pi^+\pi^+$ 
invariant mass range is consistent with the signal box.

This selection gives 6991 events in the signal box.
From a fit to the $m_{\rm{BC}}$ distribution, shown in 
Fig.~\ref{figs:fig_mode_209_dalitz_p43_data}, we find 
2159$\pm$18 of these to be background.
The $K^0_S \to \pi^+\pi^-$ contribution to the sample of events in the signal box
is easily seen as a sharp peak in the invariant $\pi^+\pi^-$ mass spectrum
shown in 
Fig.~\ref{figs:fig_mode_209_dalitz_p12_data}.
The $K^0_S$ contribution is well described by a Gaussian shape
with resolution $\sigma(m_{\pi^+\pi^-}) = 3.5$~MeV/$c^2$ both in data and
the simulation.
The number of events in the $K^0_S$ peak is 2239$\pm$77 from a fit
to a Gaussian signal plus linear background. 
Excluding $K^0_S\pi^+$ fraction and the background
leaves $\sim$2600 signal events of the $D^+\to \pi^-\pi^+\pi^+$ decay.
From these yields we calculate branching fractions,
${\mathcal B}(D^+ \to \pi^-\pi^+\pi^+) = (0.33\pm0.01)\%$ and 
${\mathcal B}(D^+ \to K^0_S\pi^+)      = (1.59\pm0.06)\%$
(statistical errors are shown only),
which are consistent with recently published CLEO-c results
${\mathcal B}(D^+ \to \pi^-\pi^+\pi^+) = (0.34\pm0.02)\%$
\cite{Blusk} and
${\mathcal B}(D^+ \to K^0_S\pi^+)      = (1.55\pm0.05\pm0.06)\%$
\cite{HadronicBF}. 
This cross-check demonstrates the quality of our simulation 
and validity of assumptions about the background level.

The presence of two $\pi^+$ mesons impose a Bose-symmetry
of the $\pi^-\pi^+\pi^+$ final state.
The Bose-symmetry when interchanging the two same sign charged pions
is explicitly accounted for in our amplitude parameterization.
We analyze events on the Dalitz plot by choosing 
$x \equiv m^2(\pi^+\pi^-)_{\rm{Low}}$  and 
$y \equiv m^2(\pi^+\pi^-)_{\rm{High}}$ as the
independent ($x$,$y$) variables.  
The third variable $z \equiv m^2(\pi^+\pi^+)$ is dependent on $x$ and $y$
through the energy-momentum balance equation.
This choice folds all
the data into the top half of the kinematically allowed region,
as is shown in Fig.~\ref{figs:pipipi_sig_DP_dots}.
The contribution from $D^+ \to K^0_S\pi^+$
is clearly seen as the narrow vertical band with 
$m(\pi^+\pi^-)_{\rm{Low}} \simeq m_{K^0_S}$.
In our Dalitz plot analysis we do not consider events in the band
$0.2 < m^2(\pi^+\pi^-)_{\rm{Low}} < 0.3$~(GeV/$c^2$)$^2$,
which is approximately ten times our $K^0_S \to \pi^+\pi^-$ mass
resolution. This leaves 4086 (signal and background) events
for our Dalitz plot analysis.


\section{Dalitz Plot Analysis}
\label{sec:dalitz_plot_analysis}


\subsection{Formalism}

This Dalitz plot analysis exploits the
techniques and formalism described in Ref.~\cite{Tim} 
that have been applied in many other CLEO analyses.
We use an unbinned maximum likelihood fit that minimizes
the sum over $N$ events: 
\begin{equation}
\label{eqn:LogL}
	\mathcal{L} = -2\sum_{n=1}^{N} \log {\cal P}(x_n,y_n),
\end{equation}
where ${\cal P}(x,y)$ is the probability density function (p.d.f.), depends on 
the event sample to be fit,
\begin{equation}
\label{eqn:PDF}
{\cal P}(x,y) = \left\{
           \begin{array}{ll}
                  \varepsilon(x,y) & {\rm for~efficiency;} \\
                  B(x,y)           & {\rm for~background;} \\
                  f_{\rm{sig}} {\mathcal N}_S |{\mathcal M}(x,y)|^2 \varepsilon(x,y) 
                + (1-f_{\rm{sig}}) {\mathcal N}_B B(x,y)
                                   & {\rm for~signal.}
            \end{array}
              \right.
\end{equation}
The shapes 
for the efficiency, $\varepsilon(x,y)$, and background, $B(x,y)$, are
explicitly $x-y$ symmetric, third order polynomial functions. 
To account for efficiency loss in the corners of the Dalitz plot,
due to low momentum tracks that are not reconstructed,
we use three multiplicative threshold functions that drop
the efficiency to zero when one of the Dalitz variables $x$, $y$, or $z$
is at their maximum values.
The background shape parameterization also
includes the non-coherent addition of three resonances 
$\rho(770)$, $f_2(1270)$, and $K^0_S$. 
The signal p.d.f. is proportional to the 
efficiency-corrected matrix element squared, $|{\mathcal M}(x,y)|^2$,
whose fraction is $f_{\rm{sig}}$. 
We estimate $f_{\rm{sig}}=0.548\pm0.013$ from the fit to the $m_{\rm{BC}}$ mass spectrum
after removing events of the $K^0_S$ contribution.
The background term has a relative $(1-f_{\rm{sig}})$ fraction.  
The signal and the background fractions are normalized separately,
$1/{\mathcal N}_S = \int |{\mathcal M}(x,y)|^2 \varepsilon(x,y) dx dy$,
$1/{\mathcal N}_B = \int B(x,y) dx dy$, which provides the overall p.d.f. normalization,
$\int {\cal P}(x,y) dx dy =1$.
The matrix element is a sum of partial amplitudes,
\begin{equation}
\label{eqn:MatrixElement}
{\mathcal M} = \sum_R c_R A_R \Omega_R F_R,
\end{equation}
where $A_R$ is a mass and spin-dependent function,
$\Omega_R$ is an angular distribution \cite{Tim}, and
$F_R$ is the Blatt-Weisskopf angular momentum barrier-penetration factor \cite{Blatt-Weisskopf}.
In our standard fit the complex factor $c_R = a_R e^{i\phi_R}$
is represented by two real numbers, an amplitude $a_R$
and a phase $\phi_R$. These are included in the list of fit parameters
and can be left to float freely or fixed.

For well established resonances, such as
$\rho(770)$,
$f_2(1270)$,
$f_0(1370)$,
$f_0(1500)$,
$f_0(1710)$, {\sl etc.},
$A_R$ is modeled with the Breit-Wigner function
\begin{equation}
\label{eqn:Breit-Wigner}
       A_R(m) = \frac{1}{m_R^2 - m^2 - i m_R \Gamma_{R}(m)},
\end{equation}
where $m$ is the $\pi^+\pi^-$ invariant mass,
$m_R$ and $\Gamma_{R}(m)$ are the resonance mass and
mass dependent width~\cite{Tim}, respectively.
The $A_R$ parameterization of the $f_0(980)$, 
whose mass, $m_{f_0}$, is close to the $K\overline{K}$ production threshold,
uses the Flatt\'e \cite{Flatte} formula
\begin{equation}
\label{eqn:Flatte}
A_{f_0(980)}(m) = \frac{1}{m_{f_0}^2 - m^2 - i [g^2_{f_0\pi\pi}        \rho_{\pi\pi}(m)
                                              + g^2_{f_0K\overline{K}} \rho_{K\overline{K}}(m)]},
\end{equation}
where $g_{f_0\pi\pi}$ and $g_{f_0K\overline{K}}$ are the $f_0(980)$ coupling constants of
the resonance to the 
$\pi\pi$ and $K\overline{K}$ final states, and 
$\rho_{ab}(m) = 2p_a/m$ is a phase space factor, calculated for the decay
products momentum, $p_a$, in the resonance rest frame.

We model a low mass $\pi\pi$ S wave, $\sigma$ or $f_0(600)$, in a number of ways.
To compare our results with E791 we try a simple spin-0 Breit-Wigner.
We also tested a
complex pole amplitude proposed in Ref.~\cite{Oller_2005}:
\begin{equation}
\label{eqn:ComplexPole}
A_{\sigma}(m) = \frac{1}{m^2_{\sigma} - m^2},
\end{equation}
where 
$m_{\sigma} = (0.47 - i 0.22)$~GeV is a pole position in the complex $s=m^2(\pi^+\pi^-)$ plane
estimated from the results of several experiments.
We also consider two
comprehensive parameterizations of the low mass $\pi\pi$ S wave.
One of them, 
suggested by J.~Schechter, is discussed in Section~\ref{sec:Schechters_Swave_in_DP_Analisis},
and its formalism is presented in Appendix~\ref{sec:appendix_schechters_amplitude}.
Another one, suggested by N.N.~Achasov, 
is discussed in Section~\ref{sec:Achasovs_Swave_in_DP_Analisis},
and its formalism is presented in Appendix~\ref{sec:achasovs_S-Wave}.


\subsection{Fits with Isobar Model}

We begin our Dalitz plot analysis
by attempting to reproduce the fit results E791~\cite{E791_Dp-pipipi}. 
Our amplitude normalization and sign conventions are different from E791. 
We therefore compare the phases and fit fractions only.
In Fit\#1 the contributions from 
$\rho(770)\pi^+$, 
$f_0(980)\pi^+$,
$f_2(1270)\pi^+$,
$f_0(1370)\pi^+$,
$\rho(1450)\pi^+$, and non-resonant intermediate states are included.
Fit\#1 gives a probability of $\simeq 0$.
We checked that the inclusion of a $\sigma \pi$ contribution, Fit \#2, 
agrees better with the data giving a fit probability of $\simeq 20\%$.
We obtain good agreement comparing our results with
Fit\#1 and Fit\#2 discussed in Ref.~\cite{E791_Dp-pipipi}.
Then, we systematically study possible contributions from all known $\pi^+\pi^-$
resonances listed in Ref.~\cite{PDG_2006}:
$\rho(770)$,
$f_2(1270)$,
$f_0(1370)$,
$\rho(1450)$,
$f_0(1500)$,
$f_0(1710)$, and
$f_0(1790)$.
We do not consider $f_2'(1525)$ due to its negligible branching fraction
to $\pi^+\pi^-$.
We assume that high mass resonances
$\rho_3(1690)$ and
$\rho(1700)$, having non-uniform angular distributions 
at the edge of the kinematically allowed region, are well 
enough represented by
$f_0(1710)$, which is a $K\bar{K}$ dominated resonance.
The asymptotic ``tails'' of other known higher mass resonances,
$f_2(1950)$,
$f_4(2050)$, 
are effectively accounted for in our fits
by the $f_0(1790)$ contribution.
We also include a unitary amplitude parametrization of the $\pi^+\pi^+$ $S$-wave
with isospin I=2 from Ref.~\cite{Achasov_PRD67_2003}.
For the $f_0(980)$ we use the Flatt\'{e} formula, Eq.~\ref{eqn:Flatte}, 
with parameters taken from the recent BES~II measurement \cite{BES_2005}.
For the $\sigma$ we switch to a complex pole amplitude, Eq.~\ref{eqn:ComplexPole},
rather than the spin-0 Breit-Wigner used by E791.

\begin{figure}
  \begin{minipage}[t]{3.0in}
    \includegraphics[width=3.0in]{\FigsDir/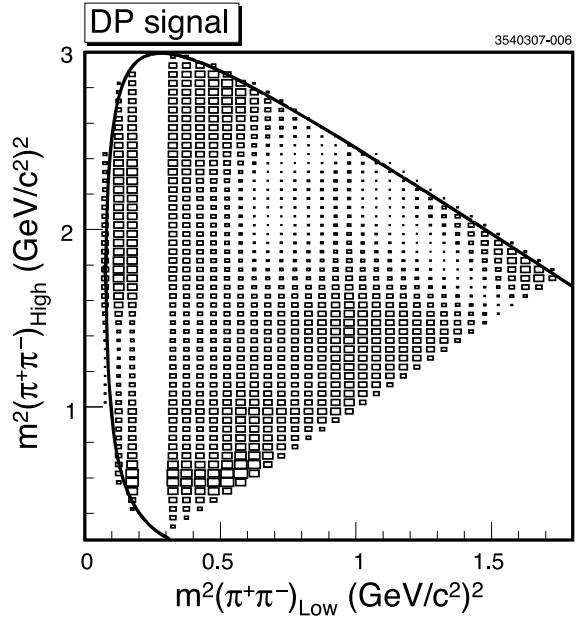}
    \caption{\label{figs:pipipi_sig_DPbox_pdf}
             The signal p.d.f. for the isobar model fit described in the text.
            }
  \end{minipage}
\end{figure}

\begin{figure}
  \begin{minipage}[t]{3.0in}
    \includegraphics[width=3.0in]{\FigsDir/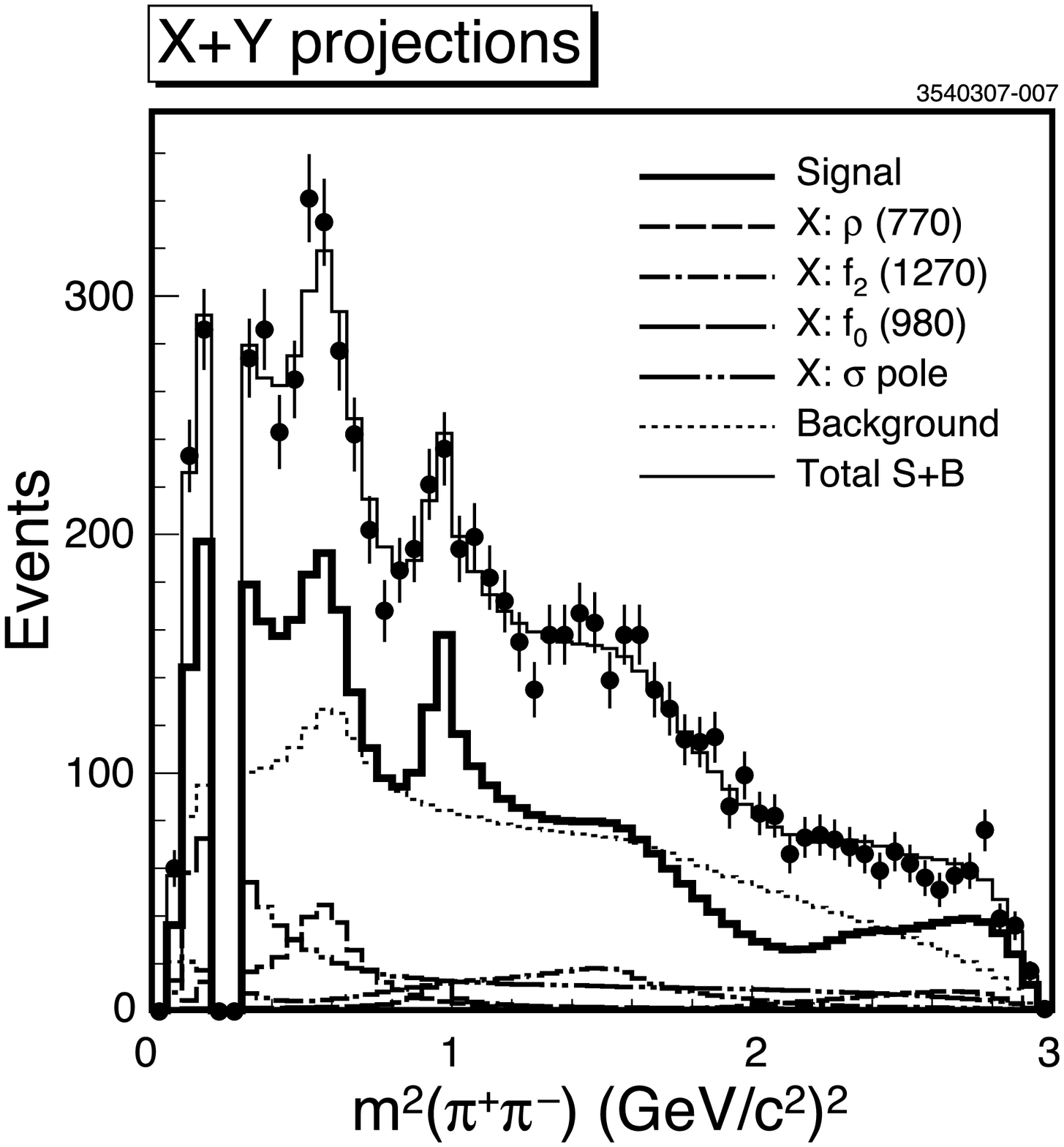}
    \caption{\label{figs:pipipi_sig_XandYproj}
             Projection of the Dalitz plot onto the $m^2(\pi^+\pi^-)$ axis 
             (two combinations per $D^+$ candidate) for CLEO-c data (points)
             and isobar model fit (histograms) showing the various components. 
            }
  \end{minipage}
  \hfill
  \begin{minipage}[t]{3.0in}
    \includegraphics[width=3.0in]{\FigsDir/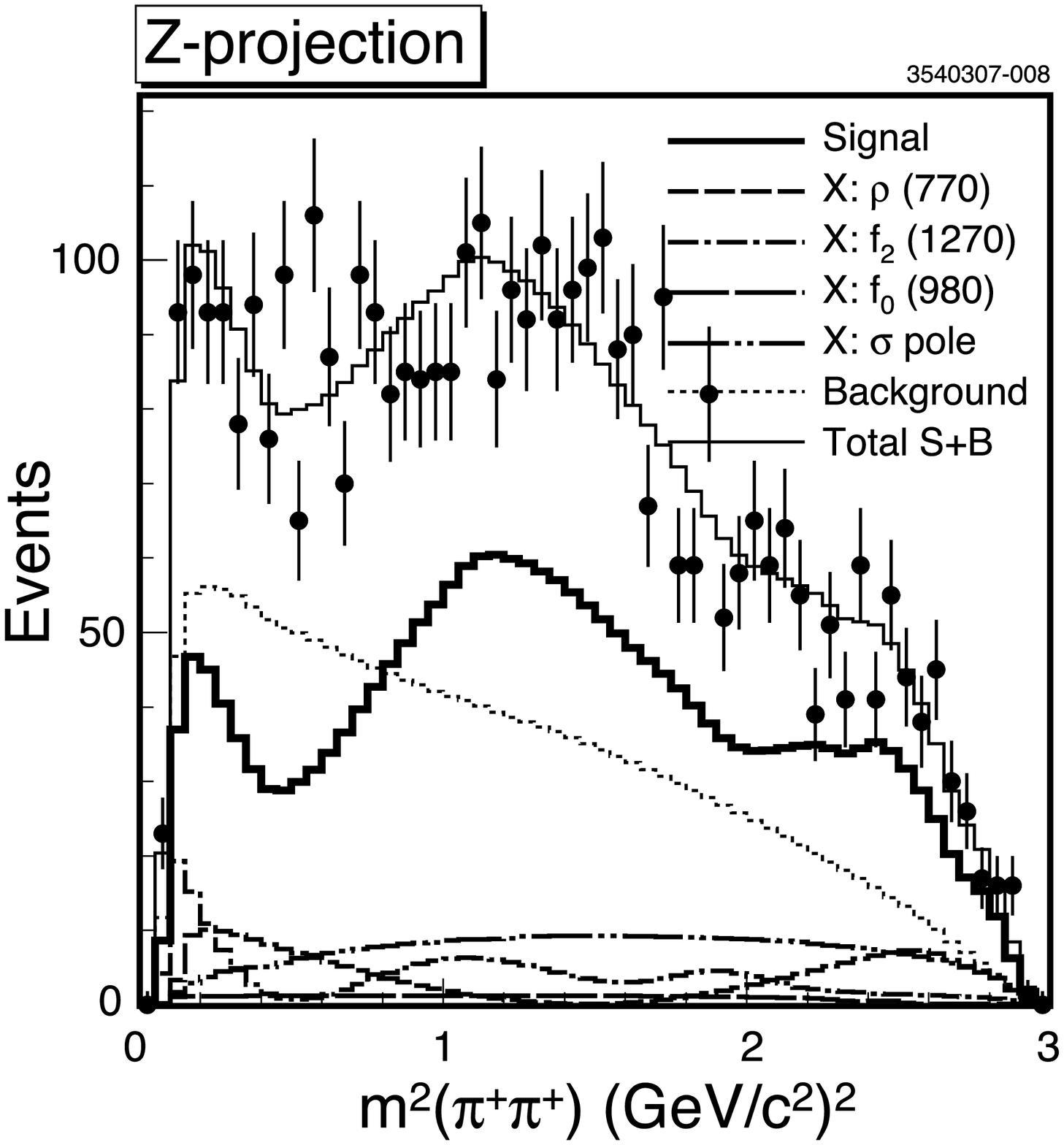}
    \caption{\label{figs:pipipi_sig_Zproj}
             Projection of the Dalitz plot onto the $m^2(\pi^+\pi^+)$ axis 
             for CLEO-c data (points)
             and isobar model fit (histograms) showing the various components. 
            }
  \end{minipage}
\end{figure}

Starting from the contributions clearly seen in our fit,
which is equivalent to Fit\#2 of E791~\cite{E791_Dp-pipipi},
we add or remove additional resonances one by one in order to improve the consistency
between the model and data.
We use Pearson's $\chi^2$ statistic criterion~\cite{PDG_2006} for adaptive bins
to calculate the probability of consistency between the p.d.f.
and the data on the Dalitz plot.  
The bins are shown in Fig.~\ref{figs:adaptive_binning_scheme}.  
We also consider the variation of the log likelihood to judge improvement.
We keep a contribution for the next
iteration if its amplitude is significant at more than three standard deviations and
the phase uncertainty is less than $30^\circ$.
Table~\ref{tab:Dp-3pi_results} shows the list of 
\begin{table}[!htb]
\caption{Results of the isobar model analysis of the $D^+\to\pi^-\pi^+\pi^+$ Dalitz plot.  
         For each contribution the relative amplitude, phase, and fit fraction
         is given.  The errors are statistical and systematic, respectively.}
\begin{tabular}{|c|c|c|c|}
\hline
\hline
Mode                &  Amplitude (a.u.)      & Phase ($^\circ$)  & Fit fraction (\%)     \\
\hline		     
$\rho(770)\pi^+$    &  1(fixed)             &  0(fixed)          &  20.0$\pm$2.3$\pm$0.9 \\
$f_0(980)\pi^+$     &  1.4$\pm$0.2$\pm$0.2  &   12$\pm$10$\pm$5  &  4.1$\pm$0.9$\pm$0.3  \\
$f_2(1270)\pi^+$    &  2.1$\pm$0.2$\pm$0.1  & --123$\pm$6$\pm$3  &  18.2$\pm$2.6$\pm$0.7 \\
$f_0(1370)\pi^+$    &  1.3$\pm$0.4$\pm$0.2  & --21$\pm$15$\pm$14 &  2.6$\pm$1.8$\pm$0.6  \\
$f_0(1500)\pi^+$    &  1.1$\pm$0.3$\pm$0.2  & --44$\pm$13$\pm$16 &  3.4$\pm$1.0$\pm$0.8  \\
$\sigma$ pole       &  3.7$\pm$0.3$\pm$0.2  &  --3$\pm$4$\pm$2   &  41.8$\pm$1.4$\pm$2.5 \\
\hline		     
\hline
\end{tabular} 
\label{tab:Dp-3pi_results} 
\end{table}
surviving contributions with their fitted amplitudes and phases, and calculated fit fractions.
The sum of all fit fractions is 90.1\%, and 
the fit probability is $\simeq$28\% for 90 degrees of freedom.
The best p.d.f. and the two projections of the Dalitz
plot and selected fit components are shown in
Figs.~\ref{figs:pipipi_sig_DPbox_pdf},
\ref{figs:pipipi_sig_XandYproj}, and~\ref{figs:pipipi_sig_Zproj}.
For contributions that are not significant we set 
upper limits at the  95\% confidence level, as shown in
Table~\ref{tab:Dp-3pi_UL}. The ``N.R.'' represents a non-resonant contribution
which is assumed to populate the Dalitz plot uniformly with a constant phase.
\begin{table}[!htb]
\caption{ Upper limit on the fit fraction, at the 95\% confidence level, 
          for contributions that we do not find significant in the
          $D^+\to\pi^-\pi^+\pi^+$ isobar model Dalitz plot analysis.}
\begin{tabular}{|c|c|}
\hline
\hline    
Mode                & Upper limit on fit fraction (\%) \\
\hline    
$\rho(1450)\pi^+$   
                    & $<$2.4        \\
		    
N.R.                
                    &$<$3.5         \\
		    
I=2 $\pi^+\pi^+$    
S wave              
                    &$<$3.7         \\
		    
$f_0(1710)\pi^+$    
                    &$<$1.6         \\
		    
$f_0(1790)\pi^+$    
                    &$<$2           \\
\hline
\hline
\end{tabular} 
\label{tab:Dp-3pi_UL} 
\end{table}

The systematic uncertainties, shown in Table~\ref{tab:Dp-3pi_results}, 
are estimated from numerous fit variations. 
We study the stability of the nominal fit results
by adding or removing degrees of freedom,
varying the list of contributions to the Dalitz plot,
changing the event selection, and varying the
efficiency and background parameterizations.
The systematic uncertainty of each fit parameter
is estimated as the quadratic sum of the mean and root mean square
values of the distribution of the changes in the parameter from
its value in the nominal fit.
For example for the poorly established resonances
$f_0(980)$, $f_0(1370)$, and $\sigma$ pole,
we allow their parameters to float and the variations
of the other fit parameters contribute to the systematic
errors. The nominal and fitted values of these
parameters are presented in Table~\ref{tab:free_fit_parameters}. 
The fit results when the parameters are allowed to float
do not vary from the nominal values by more than two standard deviations. 
\begin{table}[!htb]
\caption{Parameters for the poorly established resonances used in the nominal isobar model fit
         and their fitted values when they are allowed float.}
\begin{tabular}{|c|c|c|c|}
\hline
\hline
            & Parameter                         & Nominal Value         &  Fitted Value        \\
\hline
Signal fraction
       & $f_{\rm{sig}}$ from Eq.~\ref{eqn:PDF}  & 0.548                 & 0.552$\pm$0.020      \\
\hline
$f_0(980)$  & $m_{f_0(980)}$ (MeV/$c^2$)        & 965                   &  953$\pm$20          \\
            & $g_{f_0\pi\pi}$ (MeV/$c^2$)       & 406                   &  329$\pm$96          \\
            & $g_{f_0K\overline{K}} / g_{f_0\pi\pi}$            
                                                & 2 -- fixed            &  2 -- fixed          \\
\hline
$f_0(1370)$ & $m_{f_0(1370)}$ (MeV/$c^2$)       & 1350                  & 1259$\pm$55          \\
            & $\Gamma_{f_0(1370)}$ (MeV/$c^2$)  & 265                   &  298$\pm$21          \\
\hline
$\sigma$ pole
            & $Re(m_\sigma)$ (MeV/$c^2$)        & 470                   &  466$\pm$18          \\
            & $Im(m_\sigma)$ (MeV/$c^2$)        &--220                  & --223$\pm$28         \\
\hline
\hline
\end{tabular} 
\label{tab:free_fit_parameters}
\end{table}


\subsection{Schechter Model}
\label{sec:Schechters_Swave_in_DP_Analisis}

The isobar model drawbacks are most apparent in the
S wave $\pi^+\pi^-$ sector where wide resonances overlap
and unitarity is not fulfilled.
The model of Joseph Schechter and co-workers in 
Refs~\cite{Schechter_2001}, \cite{Schechter_2005}
is based on the meson part of the chiral invariant linear sigma model 
\cite{sigma_model} Lagrangian. 
Poles are handled using K-matrix regularization which respects unitarity by definition.
Details of the parameterization are discussed in Appendix~\ref{sec:appendix_schechters_amplitude},
and here we only summarize the meaning of the fit parameters.

In our isobar model Dalitz plot fit the $\pi^+\pi^-$ S wave
is represented by a complex pole for the $\sigma$,
the Flatt\'{e} for the $f_0(980)$ and two Breit-Wigner for the $f_0(1370)$ and $f_0(1500)$. 
Schechter's S wave amplitude, Eq.~\ref{eqn:production_amplitude} 
(Appendix~\ref{sec:appendix_schechters_amplitude}), 
parameterizes simultaneously the $\sigma$ mixed with the $f_0(980)$ 
in strong and weak interactions.
The Schechter model
describes the mixed $\sigma$ and $f_0(980)$ contributions to the Dalitz plot
with seven parameters:
the bare masses $m_\sigma$ and $m_{f_0}$;
the strong mixing angle $\psi$ between the $\sigma$ and $f_0(980)$;
the total S wave amplitude $a_{SW}$ and phase $\phi_{SW}$;
and the relative weak amplitude $a_{f_0}$ and phase $\phi_{f_0}$ of the
$f_0(980)$ with respect to the $\sigma$ amplitude.
A combination of these parameters in the model
gives the total $\pi^+\pi^-$ scattering phase, $\delta(m)$,
and an overall S wave amplitude, $A_{SW}$, for the $\sigma$ and $f_0(980)$ contributions. 
Operationally we replace the isobar $\sigma$ and $f_0(980)$
contributions by the function of Eq.~\ref{eqn:production_amplitude}
times $c_{SW} = a_{SW} e^{i\phi_{SW}}$.
The Breit-Wigner's parameterization is still used
for the $f_0(1370)$ and $f_0(1500)$.

In an initial fit \#S1, shown in Table~\ref{tab:Schechter_amp_fit_parameters},
\begin{table}[!htb]
\caption{\label{tab:Schechter_amp_fit_parameters}
         S wave amplitude parameters in the fit of the Schechter model described
         in the text to the $D^+\to\pi^-\pi^+\pi^+$ Dalitz plot.}
\begin{center}
\begin{tabular}{|l|c|c|c|}
\hline
\hline
Mode                       & \#S1        & \#S2         & \#S3        \\
\hline					                          
$m_\sigma$ (MeV/$c^2$)     & 847         & 758$\pm$36   & 745$\pm$55  \\
$m_{f_0}$ (MeV/$c^2$)      & 1300        & 1385$\pm$101 & 1221$\pm$128\\
$\psi$  ($^\circ$)         & 48.6        & 45$\pm$5     & 38$\pm$9    \\
$a_{SW}$                   & 4.1$\pm$0.2 & 3.9$\pm$0.4  & 4.5$\pm$0.6 \\ 
$\phi_{SW}$ ($^\circ$)     & 54$\pm$3    & 54$\pm$4     & 55$\pm$6    \\ 
$a_{f_0}$                  & 3.8$\pm$0.2 & 4.2$\pm$1.5  & 2.1$\pm$1.5 \\ 
$\phi_{f_0}$ ($^\circ$)    & 23$\pm$3    & 22$\pm$5     & 21$\pm$5    \\
\hline					                
$FF$(S~wave)               & 45.9$\pm$1.9& 46.4$\pm$4.8 & 43$\pm$12   \\
$\sum_i FF_i$ (\%)         & 92.1        & 90.6         & 88.3        \\
\hline          
Pearson$/N_{d.o.f.}$       & 116.3/96    & 100.4/93     & 99.6/87     \\
Probability (\%)           & 7.8         & 28.2         & 16.8        \\
$-2\sum \log L$            & 414         & 398          & 397.3       \\
\hline
\hline
\end{tabular}
\end{center}
\end{table}
we fix all amplitudes and phases to their values from our isobar model fit,
fix the S wave model parameters as in Eq.~\ref{eqn:sigma-f0_parameters},
float the S wave amplitude $a_{SW}$ and phase $\phi_{SW}$, 
and float the relative $f_0(980)$ amplitude $a_{f_0}$ and phase $\phi_{f_0}$ 
in Eq.~\ref{eqn:production_amplitude}.
This fit gives a probability of $8\%$ which indicates
the Schechter model for the S wave is an acceptable description of the data.

In a second fit, \#S2 in Table~\ref{tab:Schechter_amp_fit_parameters},
we start from the parameters obtained in \#S1
and allow the bare masses $m_\sigma$, $m_{f_0}$, 
and the strong mixing phase $\psi$ 
in Eq.~\ref{eqn:production_amplitude} to float.
This fit gives a probability of $28\%$ and 
$m_\sigma = (758\pm36)$~MeV/$c^2$, which is 
$\sim 3$ standard deviations lower than the values obtained in Ref.~\cite{Schechter_2001}, 
as also shown in our Eq.~\ref{eqn:sigma-f0_parameters}.
The mass $m_{f_0}$ and the phase $\psi$ are statistically 
consistent with the results in Ref.~\cite{Schechter_2001}.

\begin{figure}
  \begin{minipage}[t]{3.0in}
  \includegraphics[width=3.0in]{\FigsSch/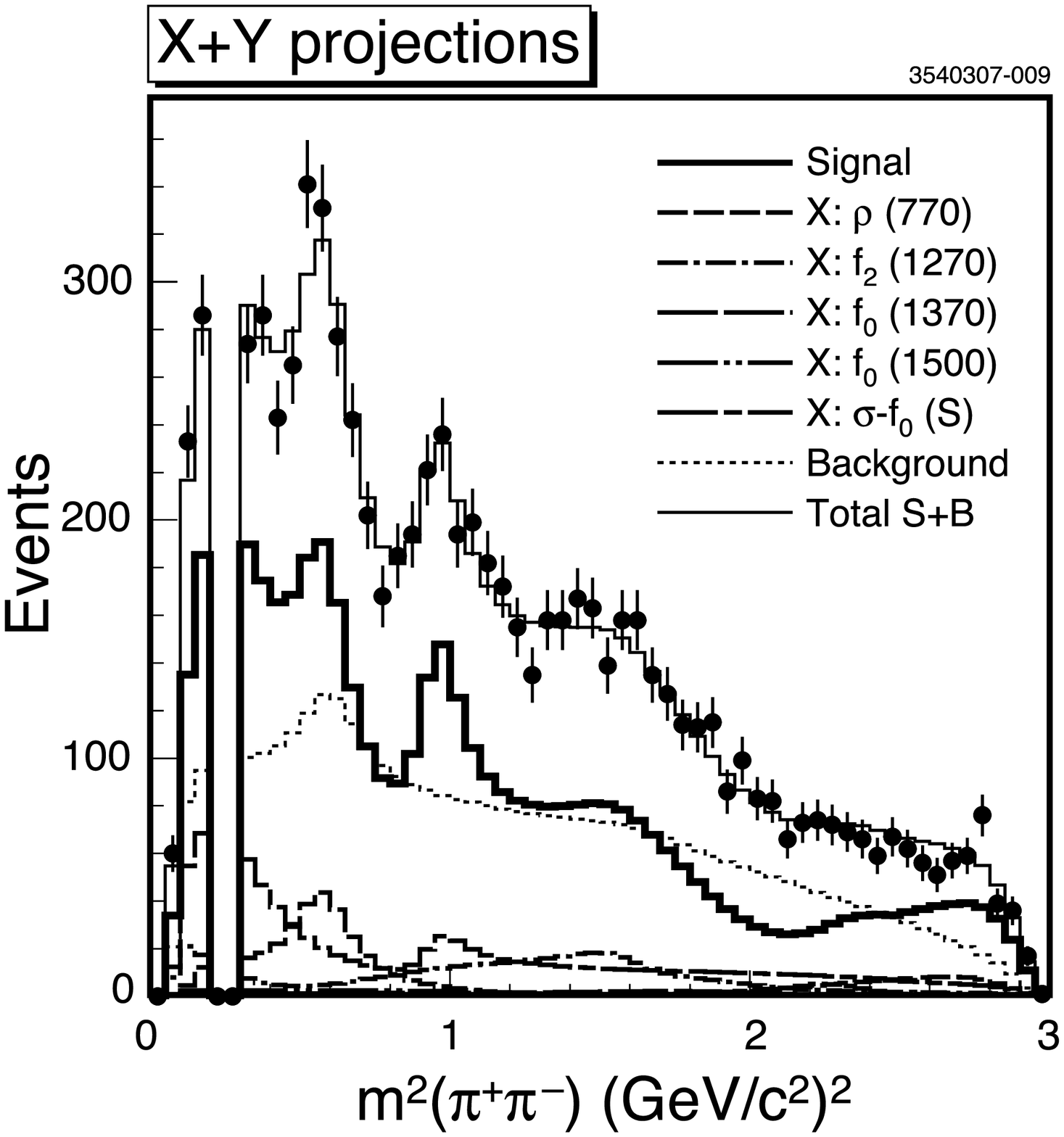}
  \caption{\label{fig:pipipi_sig_X-Y_proj_S05} 
             Projection of the Dalitz plot onto the $m^2(\pi^+\pi^-)$ axis 
             (two combinations per $D^+$ candidate) for CLEO-c data (points)
             and Schechter model fit \#S3 (histograms) showing the various components. 
          }
  \end{minipage}
  \hfill
  \begin{minipage}[t]{3.0in}
  \includegraphics[width=3.0in]{\FigsSch/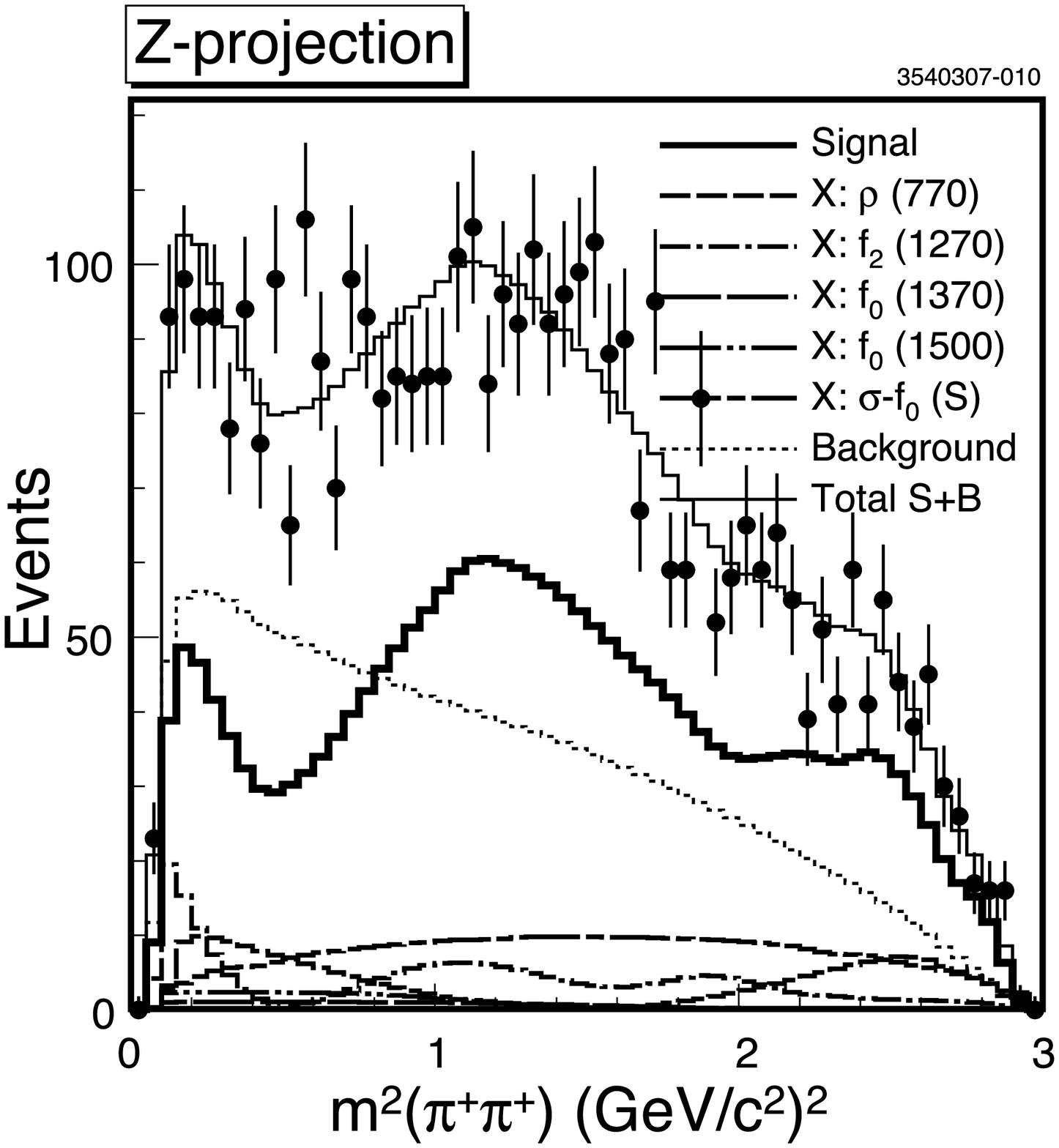}
  \caption{\label{fig:pipipi_sig_Z_proj_S05}
             Projection of the Dalitz plot onto the $m^2(\pi^+\pi^+)$ axis 
             for CLEO-c data (points)
             and Schechter model fit \#S3 (histograms) showing the various components. 
          }
  \end{minipage}
\end{figure}

\begin{figure}
  \begin{minipage}[t]{3.0in}
  \includegraphics[width=3.0in]{\FigsSch/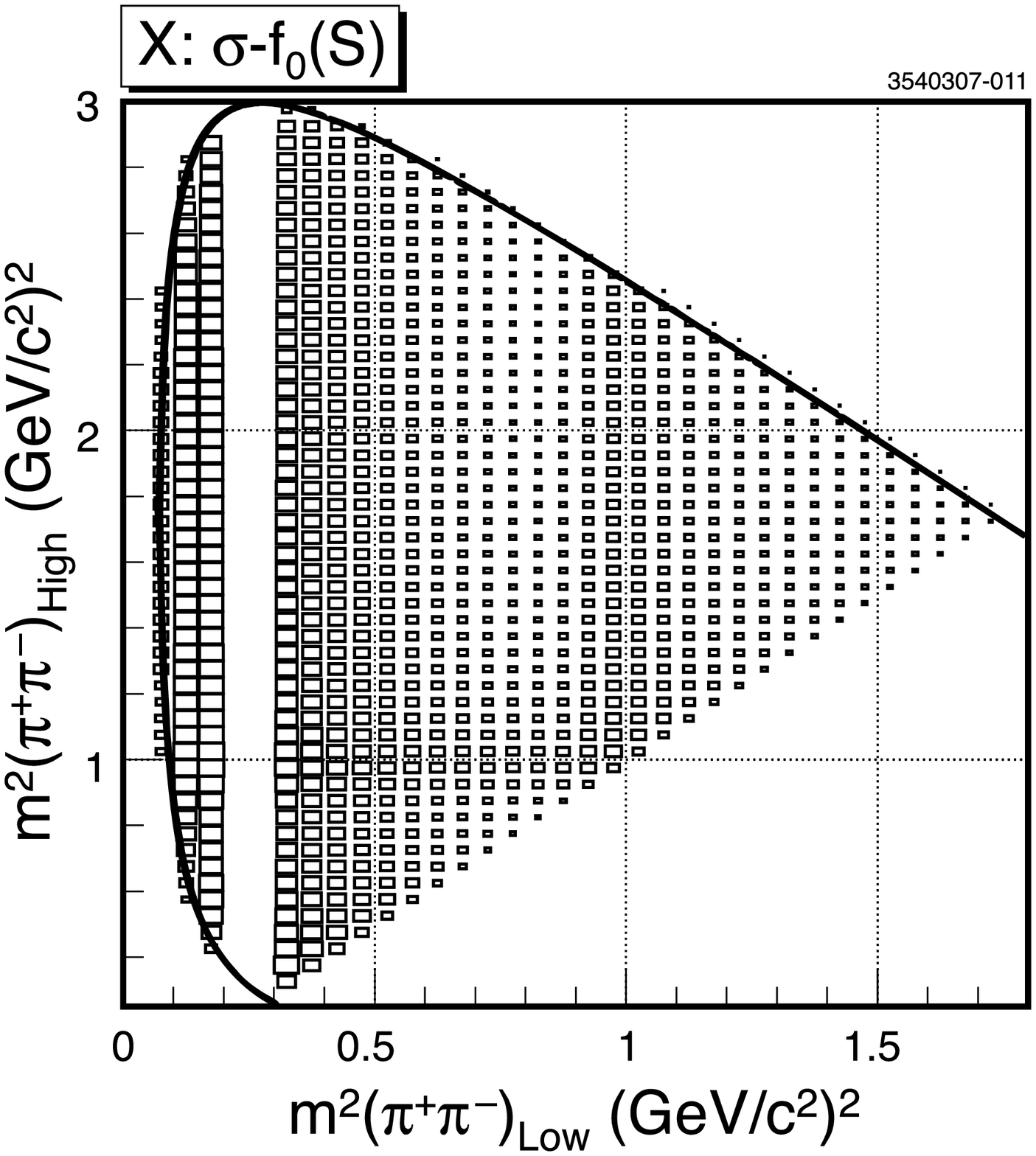}
  \caption{\label{fig:DPbox_Schechters_Swave} The isolated S wave contribution of 
                                              Schechter model fit \#S3 on the Dalitz plot.}
  \end{minipage}
  \hfill
  \begin{minipage}[t]{3.0in}
  \includegraphics[width=3.0in]{\FigsSch/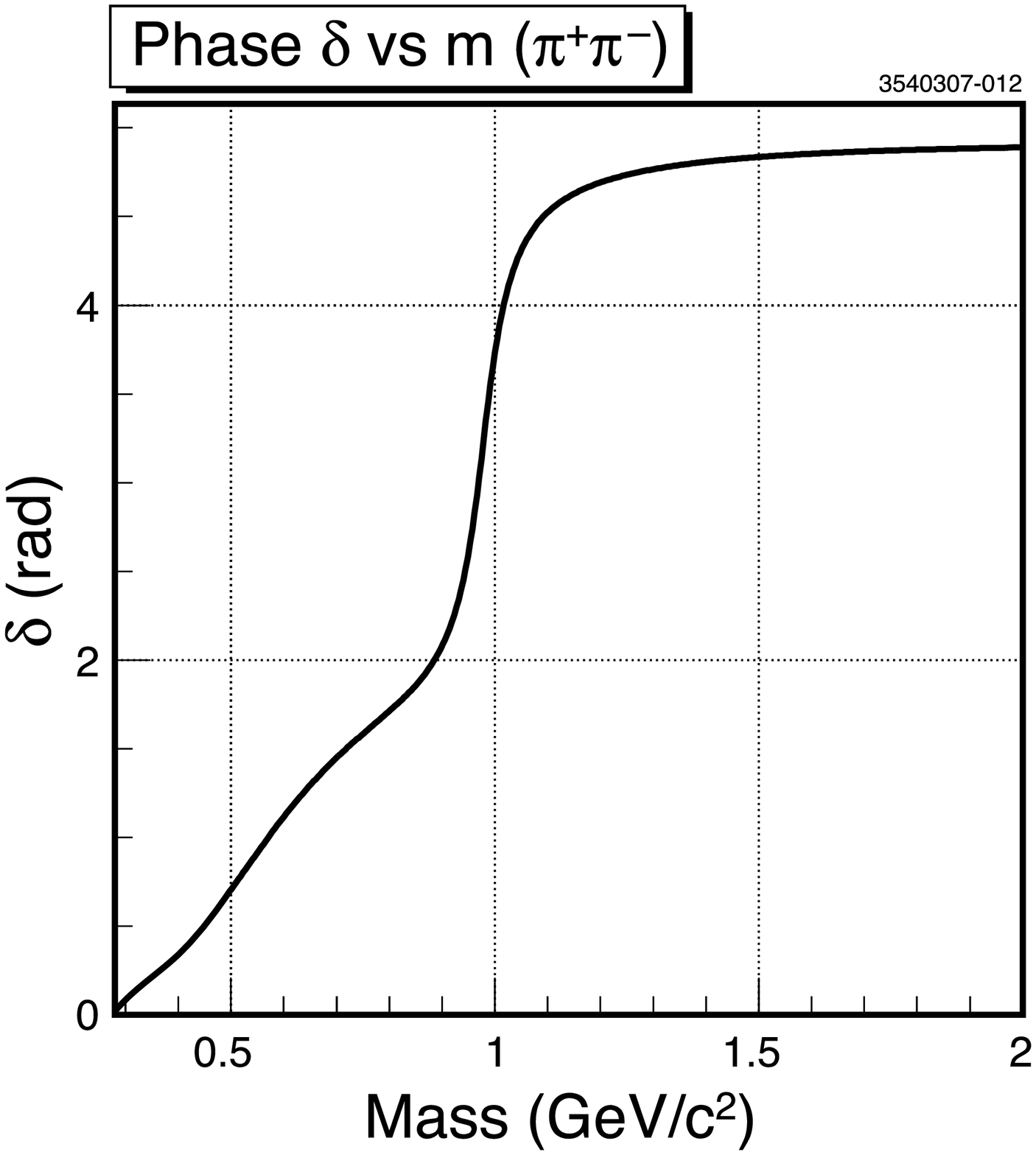}
  \caption{\label{fig:delta_S05} The $\pi^+\pi^-$ 
                                 scattering phase $\delta(m)$, Eq.~\ref{eqn:corrected_delta}, 
                                 calculated for parameters
                                 from Schechter model fit \#S3 to the 
                                 $D^+\to\pi^-\pi^+\pi^+$ Dalitz plot.}
  \end{minipage}
\end{figure}

\begin{figure}
  \includegraphics[width=5.5in]{\FigsAch/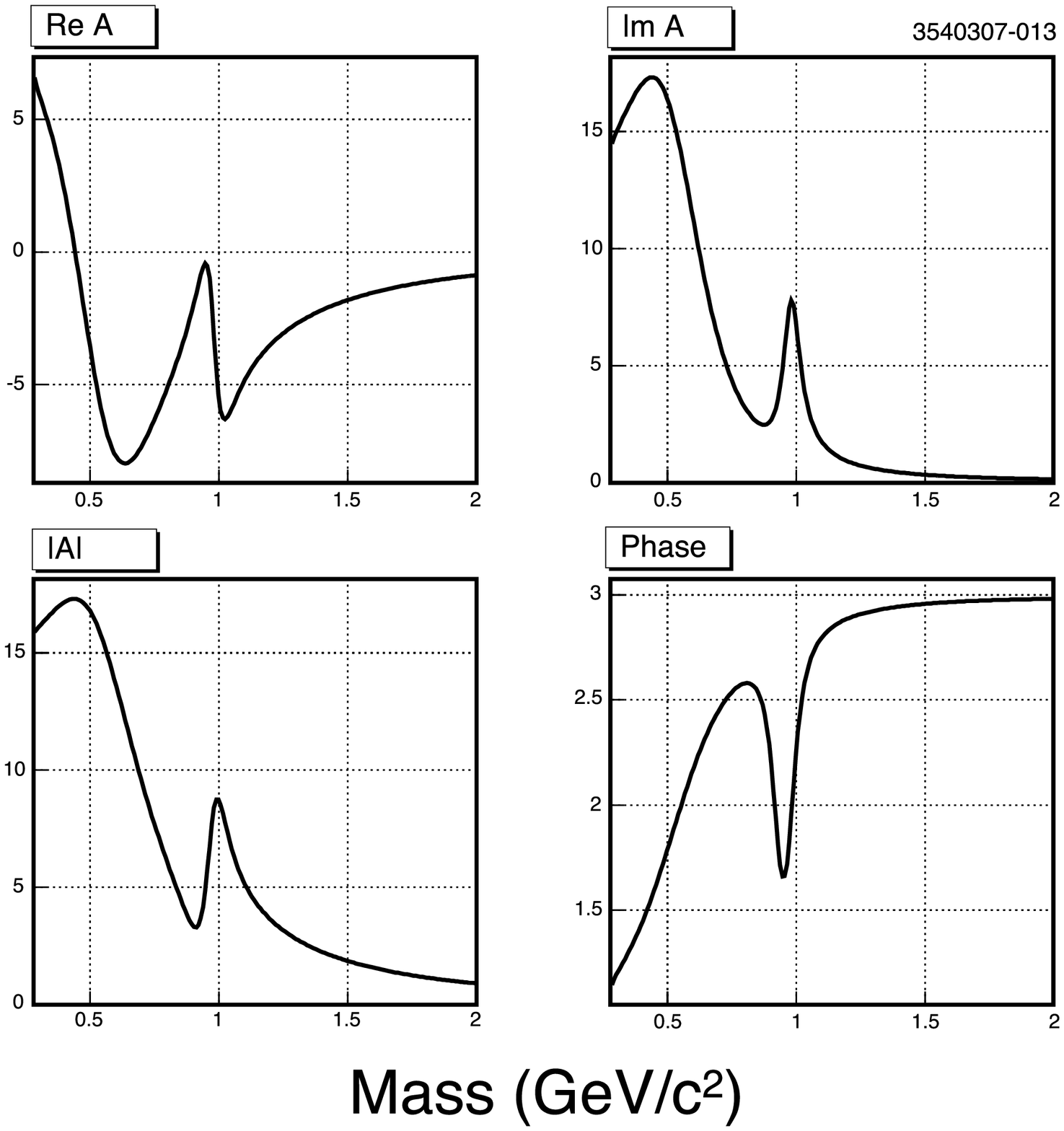}
  \caption{\label{fig:Schechter_comp_amp} 
           Complex S wave amplitude from Schechter model fit \#S3
           to the $D^+\to\pi^-\pi^+\pi^+$ Dalitz plot.
           The real and imaginary parts, the magnitude and phase are shown 
           as a function of $\pi^+\pi^-$ mass. 
          }
\end{figure}

Fits \#S1 and \#S2 are used for an initial assessment of 
the Schechter S wave parameters relative to the isobar model fit.
In a final fit, \#S3 in Table~\ref{tab:Schechter_amp_fit_parameters},
we float the Schechter S wave model parameters and
all the parameters of the other contributions.
The results of fit \#S3 are shown in 
Figs.~\ref{fig:pipipi_sig_X-Y_proj_S05} and~\ref{fig:pipipi_sig_Z_proj_S05}
in projections of the Dalitz plot.
Figure~\ref{fig:DPbox_Schechters_Swave} shows the
isolated S wave contribution to the Dalitz plot, and
Fig.~\ref{fig:delta_S05} shows the $\pi\pi$ scattering phase, $\delta(m)$,
defined in Eq.~\ref{eqn:corrected_delta}
in Appendix~\ref{sec:appendix_schechters_amplitude}.
The total signal contribution is very similar to that shown in
Fig.~\ref{figs:pipipi_sig_DPbox_pdf}.
Figure~\ref{fig:Schechter_comp_amp}
shows the complex amplitude $A_{SW}$ from Eq.~\ref{eqn:production_amplitude}
as the real and imaginary parts, the magnitude and complex phase.

Employing the Schechter model changes the fit parameters
for the non S wave contribution by less than
the systematic uncertainties in the isobar model fit.
We also note that the amplitude and fractions of $f_0(1370)$ and $f_0(1500)$
tend to be larger in the Schechter model fit.
This model gives an acceptable fit probability $\sim$17\%
when it is used to describe the $\sigma$ and $f_0(980)$ fractions in our data.
The S wave fit fraction, (43$\pm$12)\%, is consistent with a sum of fit fractions from
$\sigma$, (41.8$\pm$1.4$\pm$2.5)\%,  and $f_0(980)$, (4.1$\pm$0.9$\pm$0.3)\%
in the isobar model.
We find the Schechter S wave model parameters, 
listed in Table~\ref{tab:Schechter_amp_fit_parameters},
are consistent with the values in Ref.~\cite{Schechter_2001}.
Our data are consistent with both the isobar and Schechter models.


\subsection{Achasov Model}
\label{sec:Achasovs_Swave_in_DP_Analisis}

In Refs.~\cite{Achasov_YF32_1980}--\cite{Achasov_PRD73_2006} and references therein,
a $\pi\pi$ S wave interaction is studied for $\pi\pi \to \pi\pi$, $\pi\pi \to K\bar{K}$,
$\phi \to (f_0-\sigma)\gamma \to \pi\pi\gamma$, and $\gamma\gamma \to \pi\pi$ processes
in a manner motivated by field theory.
The $\pi\pi$ S wave production and the final state interaction
(FSI) mechanism in $D$ meson three-body decays 
have not yet been considered in the framework of this model. 
In Ref.~\cite{Achasov_D3pi} the $\pi\pi$ S wave amplitude in
$D^+\to\pi^-\pi^+\pi^+$ decay is discussed.
The developed formalism is described in Appendix~\ref{sec:achasovs_S-Wave},
and here we only summarize the meaning of the fit parameters.
The Achasov model treats the $\pi\pi$ S wave contribution to $D^+\to\pi^-\pi^+\pi^+$ 
via the sum of a number of amplitudes.  
There is a contribution
from the non-resonant, point-like $\pi^-\pi^+\pi^+$ production amplitude;
direct resonance production via the 
$D^+\to\sigma\pi^+$, 
$D^+\to f_0(980)\pi^+$;
and the rescattering terms from several intermediate states, 
$\pi^+\pi^-$, 
$\pi^0\pi^0$, and
$K\overline{K}$, to the final $\pi^+\pi^-$ state.
Our parameterization has an amplitude, $a_{D^+R\pi^+}$, and phase, $\phi_{D^+R\pi^+}$,
for the direct resonance production term, 
accounting for the $\sigma$ and $f_0$ components controlled by the coupling constants
$g_{D^+\sigma\pi^+}$ and $g_{D^+f_0\pi^+}$. 
The contributions from rescattering have 
amplitudes and phases parametrized by $a_{\rm{mode}}$ and $\phi_{\rm{mode}}$
plus a parameter from loop diagram contributions, $d_{\rm{mode}}$.
We explicitly fit for the ``mode'' = $\pi^+\pi^-$, $\pi^0\pi^0$, and $K\overline{K}$
rescattering contributions.
The contribution from non-resonant $\pi^-\pi^+\pi^+$ is also
accoonted for the relevant point-like production amplitude parameter.

We start with the parameters,
shown in Table~\ref{tab:Dp-3pi_results}, where
the $\sigma$ pole and $f_0(980)$ are replaced by the S wave amplitude from 
Eq.~\ref{eqn:Swave_amp_total}. 
We fix all resonance parameters from our isobar model fit and
float different sets of S wave parameters to assess their range.
In four fits we float the amplitude,
$a_{\rm{mode}}$, phase, $\phi_{\rm{mode}}$, and the offset parameter, $d_{\rm{mode}}$, (or 
coupling constants $g_{D^+\sigma\pi^+}$ and $g_{D^+f_0\pi^+}$
in case of direct $\sigma$ or $f_0$ meson production)
for sub-modes 
$\pi^+\pi^- \to \pi^+\pi^-$, 
$K\overline K \to \pi^+\pi^-$, 
$\pi^0\pi^0 \to \pi^+\pi^-$, or ``$DR\pi$'', respectively.  
For each of the single sub-modes we get a fit inconsistent with data.
In five fits we float $a_{\rm{mode}}$, $\phi_{\rm{mode}}$, 
$d_{\rm{mode}}$ (or $g_{D^+\sigma\pi^+}$ and $g_{D^+f_0\pi^+}$) parameters
for each combination of two sub-modes. 
All fits without the $\pi^0\pi^0 \to \pi^+\pi^-$ sub-mode
show probability of consistency with the data $\sim$10\%,
while models with the $\pi^0\pi^0 \to \pi^+\pi^-$ sub-mode
are poorly consistent with the data.
In three fits we include three or more sub-modes.
These have a consistency with the data of $\sim$10\%,
but give poor statistical significance for the amplitude parameters.
Fit \#A1 allows full freedom for all the S wave sub-modes and
gives a probability of consistency with the data of $\sim$19\%,
with 2--3 standard deviation significance for the amplitude parameters.
Its results are shown in Table~\ref{tab:DP_fits_3}.

\begin{figure}
  \begin{minipage}[t]{3.0in}
  \includegraphics[width=3.0in]{\FigsAch/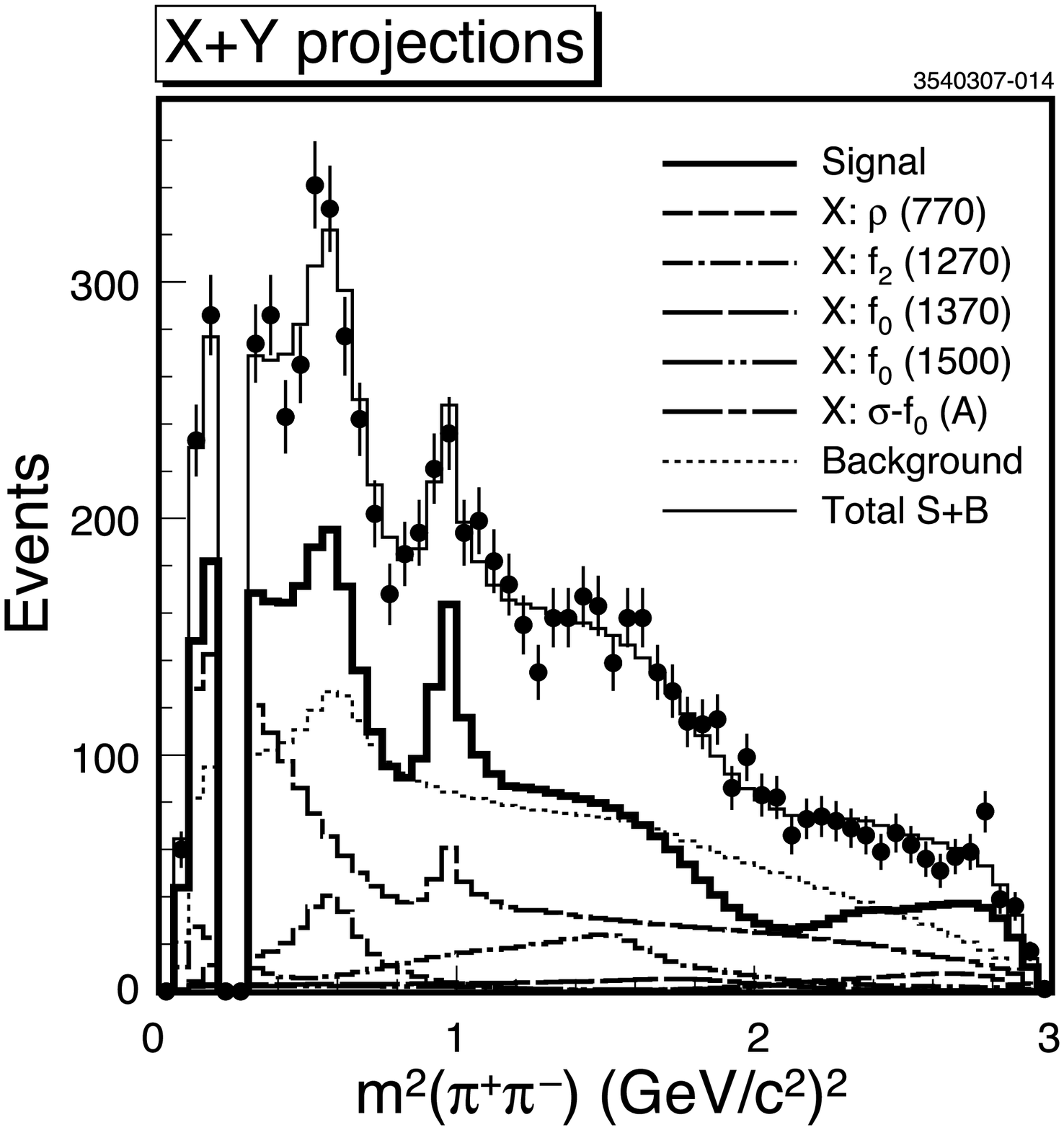}
  \caption{\label{fig:XandYproj_Achasov}
             Projection of the Dalitz plot onto the $m^2(\pi^+\pi^-)$ axis 
             (two combinations per $D^+$ candidate) for CLEO-c data (points)
             and Achasov model fit \#A2 (histograms) showing the various components. 
          }
  \end{minipage}
  \hfill
  \begin{minipage}[t]{3.0in}
  \includegraphics[width=3.0in]{\FigsAch/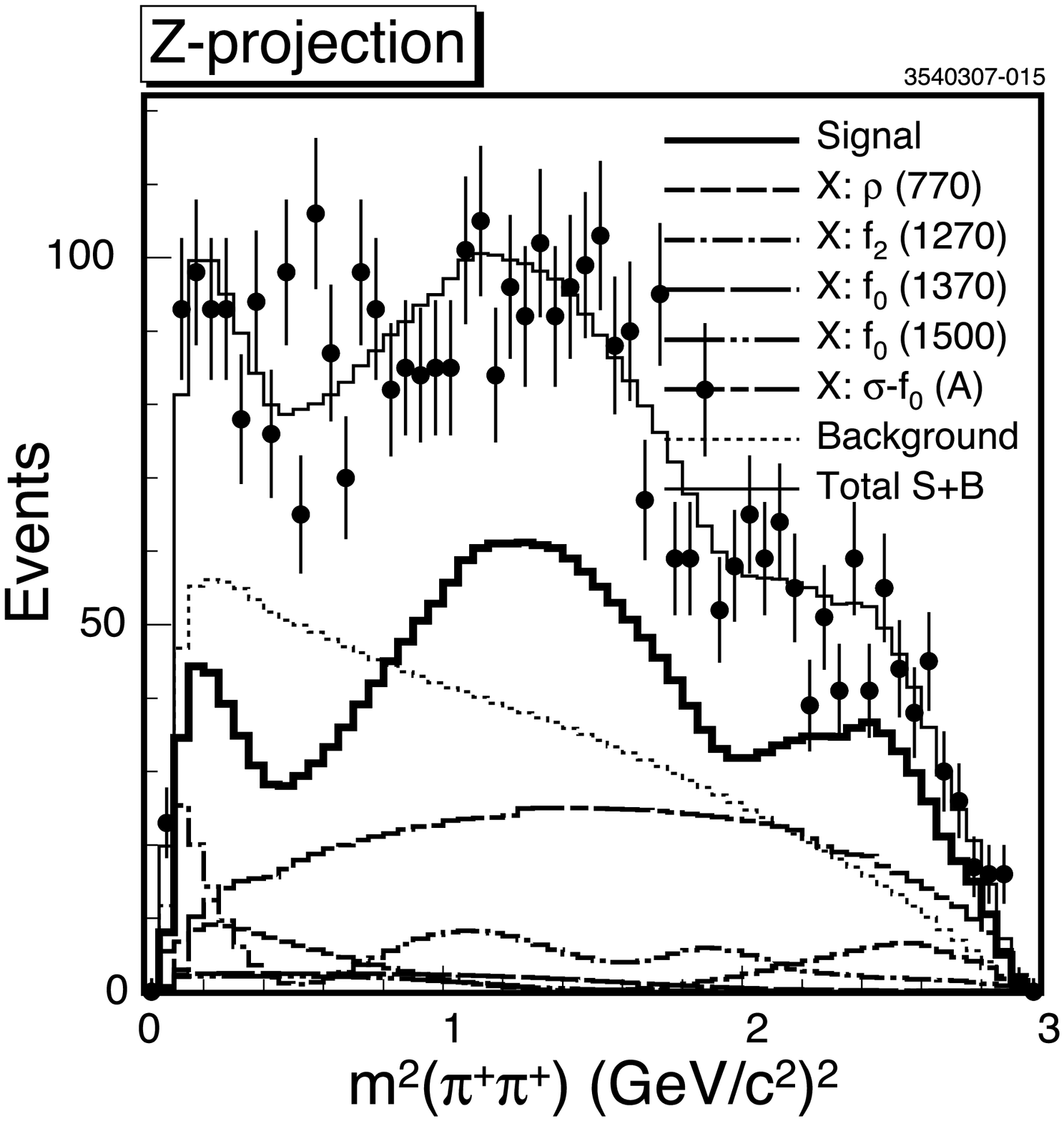}
  \caption{\label{fig:Zproj_Achasov}
             Projection of the Dalitz plot onto the $m^2(\pi^+\pi^+)$ axis 
             for CLEO-c data (points)
             and Achasov model fit \#A2 (histograms) showing the various components. 
          }
  \end{minipage}
\end{figure}

\begin{figure}
  \includegraphics[width=5.5in]{\FigsAch/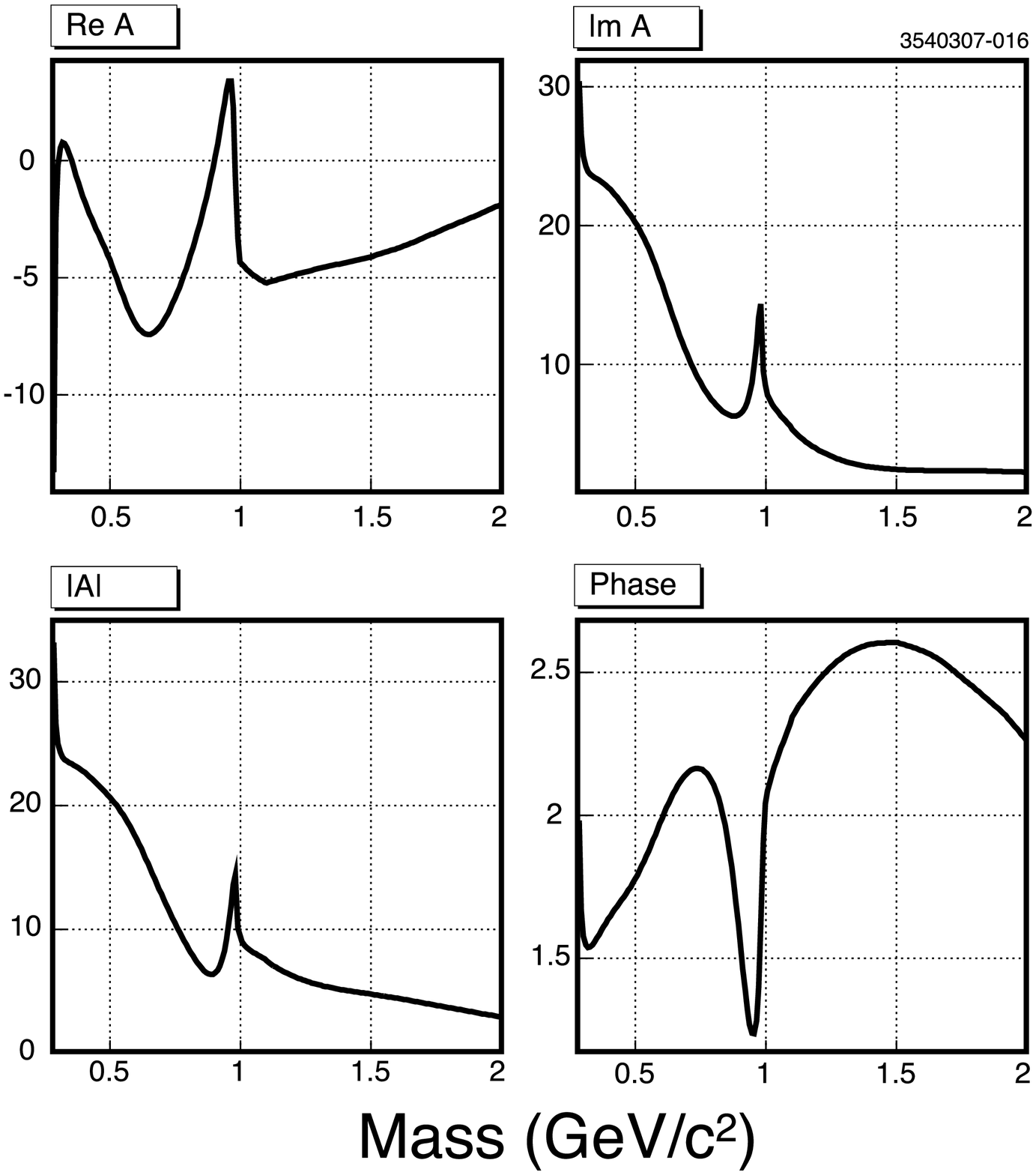}
  \caption{\label{fig:Achasov_Swave_in_fit}
           Complex S wave amplitude from Achasov model fit \#A2
           to the $D^+\to\pi^-\pi^+\pi^+$ Dalitz plot.
           The real and imaginary parts, the magnitude and phase are shown 
           as a function of $\pi^+\pi^-$ mass. 
          }
\end{figure}

We begin again with parameters of Fit \#A1
and float or set to zero amplitude the parameters of
the $f_2(1270)$, $f_0(1370)$, and $f_0(1500)$ contributions
from our isobar fit.
In Fit \#A2 we float all the S wave parameters and all
resonance parameters for the $f_2(1270)$, $f_0(1370)$, and $f_0(1500)$ contributions.
Variations of the nominal fit parameters, shown in 
Table~\ref{tab:Dp-3pi_results},
are within the range of the isobar model uncertainties.
Fit \#A3 is like Fit \#A2, but the
contributions from $f_0(1370)$ and $f_0(1500)$ scalar resonances are set to zero. 
The fit quality change from Fit \#A2 to Fit \#A3 is small.  
The S wave of the Achasov model has enough freedom 
to substitute for the contribution of the $f_0(1370)$ and $f_0(1500)$ resonances.
The results of these two fits are shown in Table~\ref{tab:DP_fits_3}.
The results of Fit \#A2 are shown in 
Figs.~\ref{fig:XandYproj_Achasov}--\ref{fig:Achasov_Swave_in_fit}
giving the Dalitz plot projections onto 
the $m^2(\pi^+\pi^-)$ and $m^2(\pi^+\pi^+)$ axes, and
the representation of the S wave complex amplitude.
Our data are consistent with the isobar, Schechter, and Achasov models.

\begin{table}[!htb]
\caption{\label{tab:DP_fits_3} Fit results for the Achasov model as described in the text.}
\small
\begin{center}
\begin{tabular}{|l|c|c|c|}
\hline
\hline
Sub-amplitude,
&\#A1         &\#A2         &\#A3         \\
parameters                   
&             &             &             \\
\hline			               	                                                           
\hline			               	                                                           
$DR\pi$                      
&             &             &             \\ 
 $a_{D^+R\pi^+}$            		 
&1--fixed     &1--fixed     &1--fixed     \\ 
 $\phi_{D^+R\pi^+}$ $(^o)$		 
&--3$\pm$32   &--66$\pm$7   &--92$\pm$13  \\
 $g_{D^+\sigma\pi^+}$       		 
&24$\pm$11    &39$\pm$8    &21$\pm$12     \\
 $g_{D^+f_0\pi^+}$          		 
&27$\pm$11    &267$\pm$24  &132$\pm$44    \\
\hline			                                                                            
 $\pi^+\pi^{\pm} \to \pi^+\pi^{\pm}$                     
&             &             &             \\
 $a_{\pi\pi}$                 
&0.25\PM0.08  &0.31\PM0.04  &0.25\PM0.07  \\
 $\phi_{\pi\pi}$ $(^o)$         
&104\PM12     &70\PM9       &93\PM9       \\
 $d_{\pi\pi}$               
&1.5\PM0.3    &2.2\PM0.2    &2.9\PM0.3    \\
\hline			      
$K\overline K \to \pi^+\pi^-$             
&             &             &             \\
$a_{K\overline K}$  
&0.56\PM0.39  &1.35\PM0.15  &1.80\PM0.40  \\
$\phi_{K\overline K}$ $(^o)$
&110\PM24     &107\PM7      &81\PM12      \\
$d_{K\overline K}$	     
&0.02\PM0.21  &0.90\PM0.09  &0.37\PM0.10  \\
\hline	
 $\pi^0\pi^0 \to \pi^+\pi^-$
&             &             &             \\
 $a_{\pi^0\pi^0}$ 
&0.13\PM0.07  &0.11\PM0.03  &  0.06\PM0.05\\
$\phi_{\pi^0\pi^0} (^o)$     
&41\PM31      &149\PM23      & 0\PM41     \\
 $d_{\pi^0\pi^0}$         
&$=d_{\pi\pi}$&$=d_{\pi\pi}$&$=d_{\pi\pi}$\\
\hline
\hline
Fit fractions (\%) 
&             &             &             \\
$\sum_i FF_i$             
&112.3        &140.4        &117.1        \\ 
$\pi^+\pi^-$, 2$\times$   
&32.1\PM9.8   &37.5\PM3.6   &34.2\PM5.3   \\ 
$\pi^+\pi^+$              
&6.1\PM5.0    &16.6\PM3.2   &9.9\PM3.0    \\ 
\hline			  
Fit goodness              
&             &             &             \\
 $Pearson/n_{d.o.f.}$
&100.7/89     &96.9/83      &106.8/87     \\
 Probability (\%)     
&18.7         &14.1         &7.3          \\
$-2\sum \log L$    
&398.6        &394.7        &405.1        \\
\hline
\hline			     
\end{tabular}
\end{center}
\end{table}


\subsection{Discussion of Models}

\begin{figure}
  \includegraphics[width=5.5in]{\FigsAch/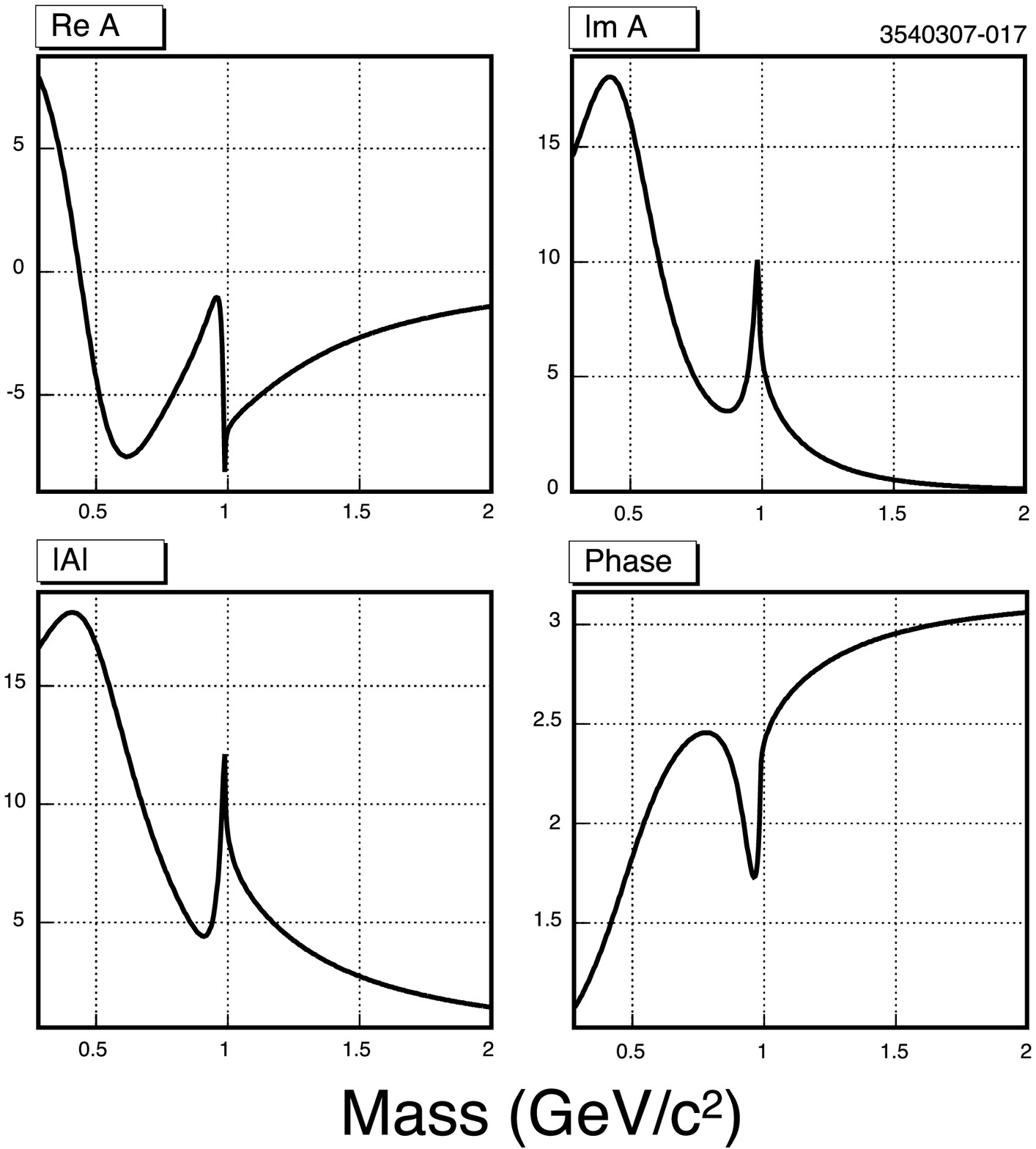}
  \caption{\label{fig:Flatte_and_ComplexPole_comp_amp}
           Complex S wave amplitude (complex pole for $\sigma$ and Flatt\'e for $f_0(980)$)
           from isobar model fit
           to the $D^+\to\pi^-\pi^+\pi^+$ Dalitz plot.
           The real and imaginary parts, the magnitude and phase are shown 
           as a function of  $\pi^+\pi^-$ mass. 
           }
\end{figure}

\begin{figure}
  \begin{minipage}[t]{3.0in}
  \includegraphics[width=3.0in]{\FigsDir/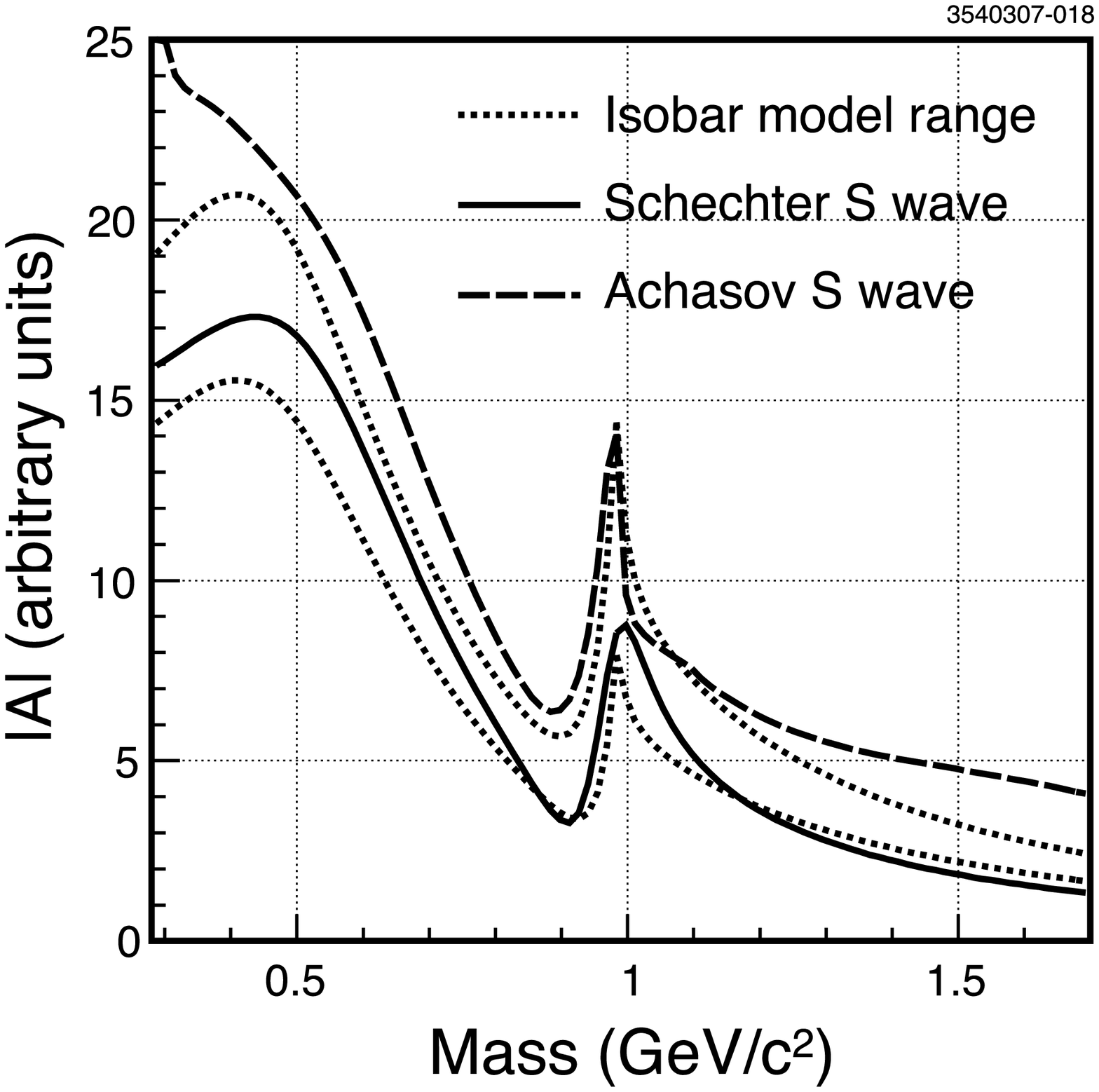}
  \caption{\label{fig:S_waves_abs_amp} 
           The $\pi^+\pi^-$ S wave absolute amplitude for different models.}
  \end{minipage}
  \hfill
  \begin{minipage}[t]{3.0in}
  \includegraphics[width=3.0in]{\FigsDir/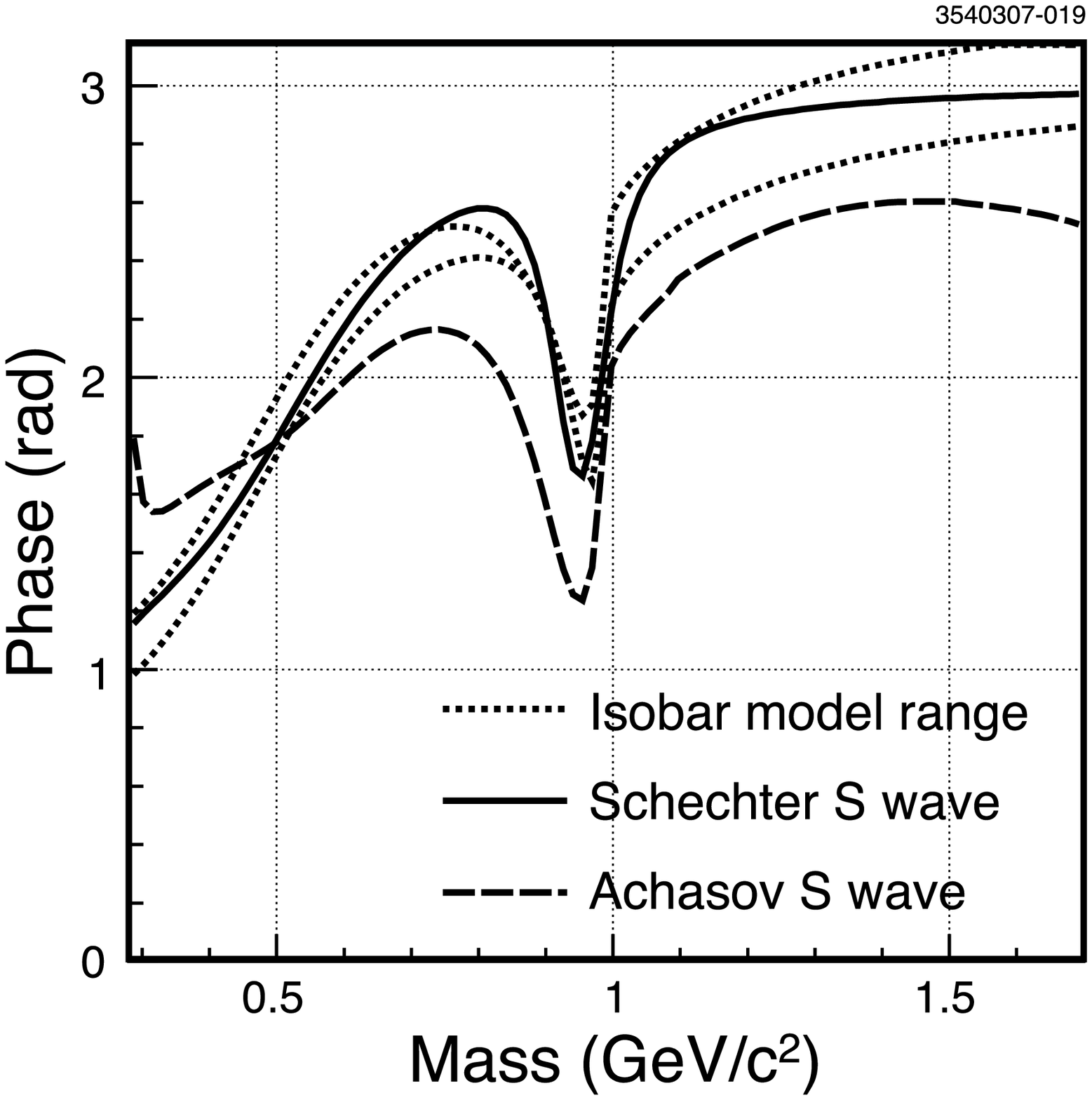}
  \caption{\label{fig:S_waves_phase} 
           The $\pi^+\pi^-$ S wave phase for different models.}
  \end{minipage}
\end{figure}

We have tested three models of the low mass $\pi^+\pi^-$ S wave in $D^+\to\pi^-\pi^+\pi^+$,
and we find little variation of the parameters describing non S wave contributions.
The fit gives similar S wave contributions for all three models.
We show this by plotting the relevant complex functions describing the S wave.
Figure~\ref{fig:Flatte_and_ComplexPole_comp_amp} shows the Flatt\'e and the complex-pole 
parameterizations for $f_0(980)$ and $\sigma$, respectively, 
for our isobar model fit to the data.
Figure~\ref{fig:Schechter_comp_amp} shows the results of the Schechter model fit, and
Fig.~\ref{fig:Achasov_Swave_in_fit} shows the results of the Achasov model fit.
In Figs.~\ref{fig:S_waves_abs_amp},~\ref{fig:S_waves_phase} we compare
the $\pi\pi$ S wave amplitude and phase in the 
accessible mass region from threshold to 1.7~GeV/$c^2$ for these three models.
The solid curve corresponds to the Schechter model fit to our Dalitz plot,
the dashed curve is for Achasov model fit, and the $\pm 1\sigma$ 
of the amplitude and phase parameters
range of the S wave contribution in the isobar model is indicated by the two dotted curves.
The S wave shapes are quite similar up to the interplay with other resonances, 
and with the data set we have in hand
we are not sensitive to the details of the S wave parameterization.


\section{Summary}
\label{sec:summary}

Using a sample of 0.78~million 
$e^+e^- \to \psi(3770) \to D^+D^-$ events collected in the CLEO-c experiment,
we performed a Dalitz plot analysis of the $D^+\to\pi^-\pi^+\pi^+$ decay.
Our nominal results, obtained within the framework of the isobar model and 
shown in Table~\ref{tab:Dp-3pi_results},
reinforce the previous conclusion \cite{E791_Dp-pipipi}, \cite{FOCUS_Dp-pipipi}
that a sizable $\sigma\pi^+$ component is required,
in addition to other intermediate states
$\rho(770)\pi^+$,
$f_2(1270)\pi^+$,
$f_0(1370)\pi^+$,
$f_0(1500)\pi^+$, and
$f_0(980)\pi^+$, in order to describe the $D^+\to\pi^-\pi^+\pi^+$ decay.
The systematic uncertainties are estimated by varying the fit parameters from 
their nominal values.
We also show in Table~\ref{tab:free_fit_parameters}
a set of optimal parameters for the $\sigma$, $f_0(980)$, and $f_0(1370)$ resonances
based on our isobar model fit to the $D^+\to\pi^-\pi^+\pi^+$ Dalitz plot.
Limits on contributions from
$\rho(1450)\pi^+$,
non-resonant,               
I=2 $\pi^+\pi^+$ S wave,             
$f_0(1710)\pi^+$, and   
$f_0(1790)\pi^+$, shown in Table~\ref{tab:Dp-3pi_UL}, are set at 95\% confidence level.

We tested other models of the low mass $\pi\pi$ S wave
contributions and in each case obtain optimal parameters. 
In Table~\ref{tab:Schechter_amp_fit_parameters}
we summarize results for the model suggested by J.~Schechter and co-workers
\cite{Schechter_2005}, \cite{Schechter_2001}.
All fits for this model show consistent values for the parameters.
We also apply the S wave model suggested by N.N.~Achasov {\it et al.} \cite{Achasov_D3pi}. 
This model has more freedom in sub-modes than we are confidently able to define with our data.
Possible solutions are presented in Table~\ref{tab:DP_fits_3}. 
Further progress with this model can be achieved if several $D$ meson decay modes
with higher statistics are analyzed simultaneously.

\begin{table}
\caption{A comparison of the observed fit fractions ($FF$) in \% in the three
         models of $D^+ \to \pi^-\pi^+\pi^+$.  For the ``Isobar'' column, the
         ``Low S wave $\pi^+$'' entry is the sum of the two entries above.
         }
\begin{tabular}{c|c|c|c}
\hline
\hline
Mode                & Isobar         & Schechter \#S3 & Achasov \#A2   \\ \hline
$\sigma \pi^+$      & $41.8 \pm 2.9$ &                &                \\ 
$f_0(980) \pi^+$    & $ 4.1 \pm 0.9$ &                &                \\ \hline
Low S wave $\pi^+$  & $45.9 \pm 3.0$ & $43.4 \pm 11.8$& $75   \pm 7$   \\
$f_0(1370) \pi^+$   & $ 2.6 \pm 1.9$ & $2.6  \pm 1.7$ & $3.2  \pm 0.7$ \\
$f_0(1500) \pi^+$   & $ 3.4 \pm 1.3$ & $4.3  \pm 2.4$ & $4.0  \pm 0.8$ \\ 
$\rho^0(770) \pi^+$ & $20.0 \pm 2.5$ & $19.6 \pm 7.4$ & $18.4 \pm 4.0$ \\
$f_2(1270) \pi^+$   & $18.2 \pm 2.7$ & $18.4 \pm 7.4$ & $23.2 \pm 5.0$ \\
I=2 $\pi^+\pi^+$ S wave &            &                & $16.6 \pm 3.2$ \\ 
\hline
$\sum_i FF_i$       & 90.1           & 88.3           & 140.4          \\ 
\hline
\hline
\end{tabular}
\label{tab:compare2}
\end{table}

For all $\pi\pi$ S wave models we find that their fit fraction exceeds 50\%, and 
confirm results of previous experiments of a significant contribution from 
a low mass $\pi^+\pi^-$ S wave in the $D^+ \to \pi^-\pi^+\pi^+$ decay. 
Table~\ref{tab:compare2} compares the fit fractions from the fits to the
three models described above.  
The S wave fit fraction in Achasov model is three standard deviation
larger than in the Isobar and Schechter model. 
The sum of all fit fractions is also larger in Achasov model, that indicates
on difference in interference terms. 
The fit fractions for sub-modes are consistent between these three models.
Figures~\ref{fig:S_waves_abs_amp} and \ref{fig:S_waves_phase}
compare the amplitude and phase, respectively, for the $\pi^+\pi^-$ S wave
contribution we have found in the three considered models.
With our given data sample all three S wave parameterizations adequately describe 
the $D^+ \to \pi^-\pi^+\pi^+$ Dalitz plot.


\section{Acknowledgments}
We thank Joseph Schechter, Amir Fariborz, and Nikolay Achasov
for stimulating discussions and significant help in application of
the low mass $\pi\pi$ S wave models.
We gratefully acknowledge the effort of the CESR staff
in providing us with excellent luminosity and running conditions.
D.~Cronin-Hennessy and A.~Ryd thank the A.P.~Sloan Foundation.
This work was supported by the National Science Foundation,
the U.S. Department of Energy, and
the Natural Sciences and Engineering Research Council of Canada.


\section{Appendix: Alternative models of the $\pi\pi$ S wave}
\label{sec:appendix}
\subsection{Formalism of the $\pi\pi$ S wave suggested by J.~Schechter}
\label{sec:appendix_schechters_amplitude}

A tree level $\pi\pi \to \pi\pi$ scattering
amplitude for two resonances $\sigma$ and $\tilde\sigma$ 
strongly-mixed with phase $\psi$ is given in 
Eq.~3.2 of Ref.~\cite{Schechter_2001}:
\begin{equation}
\label{eqn:strong_mixed_tree_amplitude}
T^0_{0~tree}=\cos^2\psi \bigg[ \alpha(s) + \frac{\beta(s)}{m^2_{\sigma} - s} \bigg]
            +\sin^2\psi \bigg[ \tilde{\alpha}(s) + \frac{\tilde{\beta}(s)}{m^2_{\tilde\sigma} - s} \bigg],
\end{equation}
where
\begin{equation}
\label{eqn:alpha}
  \alpha(s) = \frac{\sqrt{1-\frac{4m_\pi^2}{s}}}{16\pi F_\pi^2}(m_\sigma^2-m_\pi^2)
              \bigg[-5+2\frac{m_\sigma^2-m_\pi^2}{s-4m_\pi^2}
                    \ln\bigg(\frac{m_\sigma^2+s-4m_\pi^2}{m_\sigma^2}\bigg)\bigg],
\end{equation}
\begin{equation}
\label{eqn:beta}
  \beta(s) = \frac{3\sqrt{1-\frac{4m_\pi^2}{s}}}{16\pi F_\pi^2}(m_\sigma^2-m_\pi^2)^2,
\end{equation}
$s$ is the $\pi^+\pi^-$ invariant mass squared,
$m_\pi$ and $F_\pi$=0.131~GeV are the pion mass and the decay constant,
$m_\sigma$ and $m_{\tilde\sigma}$ are the bare masses of two scalar resonances,
and $\psi$ is a strong mixing angle. 
We use the original notation of Ref.~\cite{Schechter_2001},
the tilde is used for all parameters relating to
the second scalar resonance, $\tilde\sigma$,
which in our case is associated with $f_0(980)$.
Equation~\ref{eqn:strong_mixed_tree_amplitude} can be re-written as
\begin{equation}
T^0_{0~tree} = A + \frac{B}{P},
\end{equation}
where 
\begin{equation}
A = \alpha\cos^2\psi + \tilde{\alpha} \sin^2\psi,
\end{equation}
\begin{equation}
B = \beta        \cdot (m^2_{\tilde\sigma} - s) \cos^2\psi 
  + \tilde{\beta}\cdot (m^2_{      \sigma} - s) \sin^2\psi,
\end{equation}
\begin{equation}
\label{eqn:numerator_P}
P = (m^2_{\sigma} - s)(m^2_{\tilde\sigma} - s).
\end{equation}
According to the Dyson equation for the $\pi\pi$ scattering, 
Eq.~3.3 from Ref.~\cite{Schechter_2001} gives an expression for a 
total scattering amplitude through the tree amplitude:
\begin{equation}
\label{eqn:scattering_amplitude}
T^0_0(s) = \frac{T^0_{0~tree}(s)}{1-iT^0_{0~tree}(s)}.
\end{equation}
The scattering amplitude is a complex number, 
$T^0_0(s) = | T^0_0(s)| e^{i\delta(s)}$, then
the tree amplitude can be associated with the 
tangent of the scattering phase, 
\begin{equation}
T^0_{0~tree}(s) = \tan \delta(s),
\end{equation}
and we get an expression for $\cos \delta$:
\begin{equation}
\label{eqn:cos_delta_Schechter}
\cos \delta = \frac{1}{\sqrt{1 + \tan^2 \delta}} 
            = \frac{P}{\sqrt{P^2 + (P\cdot A+B)^2}}.
\end{equation}
Expression 
$\delta(s) = \arctan\Big(T^0_{0~tree}(s)\Big)$ 
defines a scattering phase in the range 
$[-\frac{\pi}{2},\frac{\pi}{2}]$.
This phase $\delta(s)$ has two discontinuities at 
$s=m^2_{\sigma}$ and $s=m^2_{f_0}$
for parameters taken from
Ref.~\cite{Schechter_2001},
\begin{equation}
\label{eqn:sigma-f0_parameters}
m_{\sigma} = 0.847~\rm{GeV}/c^2,~~~
m_{f_0} = 1.30~\rm{GeV}/c^2,~~~
\psi = 48.6^\circ.
\end{equation}
In order to remove discontinuities we add a phase-shift $+\pi$ above each bare mass:
\begin{equation}
\label{eqn:corrected_delta}
   \delta(s) = \arctan\Big(T^0_{0~tree}(s)\Big) + \pi \Big(\theta(s-m^2_{\sigma})  
                                                         + \theta(s-m^2_{f_0}) \Big),
\end{equation}
where $\theta(x)$ is a step function,
that makes the phase smooth, as shown in Fig.~\ref{fig:delta_S05}.

In this model the
production amplitude is obtained from the total scattering amplitude,
Eq.~\ref{eqn:scattering_amplitude}, by replacing the first tree level 
$\pi\pi \to \pi\pi$ scattering diagram amplitude,
$T^0_{0~tree}$, by the resonance propagator $(m^2_{\sigma} - s)^{-1}$
with the coupling constant $g_{\sigma\pi\pi}$
and keeping the proper re-scattering amplitude, represented by the ``bubble sum'' 
factor $(1-iT^0_{0~tree})^{-1}$:
\begin{equation}
\label{eqn:production_amplitude_for_one_resonance}
  {\mathcal A_\sigma} = \frac{g_{\sigma\pi\pi}}{m^2_{\sigma} - s} \cdot \frac{1}{1-iT^0_{0~tree}} 
                      = \frac{g_{\sigma\pi\pi}}{m^2_{\sigma} - s} \cdot \cos\delta \cdot e^{i\delta}.
\end{equation}
Extending Eq.~\ref{eqn:production_amplitude_for_one_resonance} 
(Eq.~15 from Ref.~\cite{Schechter_2005})
for the case of two resonances $\sigma$ and $f_0(980)$
we get the total production amplitude with relative weak interaction mixing factor 
$a_{f_0} e^{i\phi_{f_0}}$
\begin{equation}
\label{eqn:production_amplitude}
      A_{SW} = {\mathcal A}_\sigma + {\mathcal A}_{f_0} 
             = \cos\delta \cdot e^{i\delta} \bigg[ \frac{1}{m^2_{\sigma} - s}
             + \frac{a_{f_0} e^{i\phi_{f_0}}}{m^2_{f_0} - s} \bigg].
\end{equation}
Note, that Eq.~\ref{eqn:production_amplitude} does not contain singular terms because
both poles are contracted into the $P$ factor from $\cos \delta$, 
Eq.~\ref{eqn:numerator_P} and \ref{eqn:cos_delta_Schechter}. 
For the first iteration we set
\begin{equation}
\label{eqn:f0_weak_parameters}
a_{f_0}=1,~~~
\phi_{f_0} = 0^\circ.
\end{equation}

It should be noted, that in the frame of this model, 
$\sigma$ is a scalar $\pi\pi$ resonance which has a bare mass $m_\sigma$ as a parameter.
The bare mass does not coincide with a peak position as in case of Breit-Wigner,
that is clearly seen in Eq.~\ref{eqn:sigma-f0_parameters} for the mass of $f_0(980)$.
This simple model does not take in to account that the scalar resonances may have 
other decay modes, coupled channels. For example, it is well known that 
$f_0(980)$ has a $K\bar{K}$ decay mode with a mass dependent rate as large as $\sim$20\%. 
Presumably, this amplitude, obtained from the chiral Lagrangian,
works well in the region close to the production threshold. 
In the case of SU(3) symmetry it accounts for the two low mass resonances $\sigma$ and $f_0(980)$. 
Other higher mass resonances such as the $f_0(1370)$ and $f_0(1500)$ are not taken into account.
These issues restrict the precision and limit the application 
of this model.

\subsection{Formalism of the $\pi\pi$ S wave suggested by N.N.~Achasov}
\label{sec:achasovs_S-Wave}

\subsubsection{$D^+\to\pi^-\pi^+\pi^+$ total amplitude}

In this section we summarize a suggested formalism \cite{Achasov_D3pi} for a parameterization
of the $\pi\pi$ scalar amplitude in the $D^+\to\pi^-\pi^+\pi^+$ decay, and
present the details of our implementation in the Dalitz plot fitter with some relevant
cross-checks.
For the $D^+\to\pi^-\pi^+\pi^+$ decay Ref.\cite{Achasov_D3pi} suggests 
the use of a $\pi\pi$ S wave amplitude that is a superposition
\footnotesize
\begin{equation}
\label{eqn:amp_D3pi}
\begin{array}{l}
  A(D^+\to\pi^+_1\pi^+_2\pi^-) = A^{pl}(D^+\to\pi^+_1\pi^+_2\pi^-)           
\\
  +B[D^+\to\pi^+_1(\pi^+_2\pi^-\to\pi^+_2\pi^-)\to\pi^+_1\pi^+_2\pi^-]
  +B[1 \leftrightarrow 2]       
\\
  +E[D^+\to(\pi^+_1\pi^+_2\to\pi^+_1\pi^+_2)\pi^-\to\pi^+_1\pi^+_2\pi^-]
\\
  +F[D^+\to\pi^+_1(\sigma + f_0)\to\pi^+_1\pi^+_2\pi^-]
  +F[1 \leftrightarrow 2]                      
\\
  +\bar{B}[D^+\to\pi^+_1(\pi^0\pi^0\to\pi^+_2\pi^-)\to\pi^+_1\pi^+_2\pi^-]
  +\bar{B}[1 \leftrightarrow 2]   
\\
  +C[D^+\to\pi^+_1(K^+K^-\to\pi^+_2\pi^-)\to\pi^+_1\pi^+_2\pi^-]
  +C[1 \leftrightarrow 2] 
\\
  +\bar{C}[D^+\to\pi^+_1(K^0\bar{K}^0\to\pi^+_2\pi^-)\to\pi^+_1\pi^+_2\pi^-]
  +\bar{C}[1 \leftrightarrow 2].
\end{array}
\end{equation}
\normalsize
of a point-like, $A^{pl}$, direct resonance, $F$,
and non-resonant production terms, $B$, $C$, $E$, followed by the re-scattering 
in to the $\pi\pi$ final state.
Here we list the definitions of all the sub-amplitudes in Eq.~\ref{eqn:amp_D3pi}. \\
The point-like $D^+\to\pi^+_1\pi^+_2\pi^-$ amplitude is associated with a constant $a$:
\begin{equation}
\label{eqn:Apl}
A^{pl}(D^+\to\pi^+_1\pi^+_2\pi^-) = 16 \pi a.
\end{equation}
After the point-like production one would expect 
$\pi^+_1\pi^-\to\pi^+_1\pi^-$ and $\pi^+_2\pi^-\to\pi^+_2\pi^-$ scattering,
which we parametrize as a mass dependent amplitude
\[ \hspace*{-5cm}
B[D^+\to\pi^+_1(\pi^+_2\pi^-\to\pi^+_2\pi^-)\to\pi^+_1\pi^+_2\pi^-, m=m_{\pi^+_2\pi^-}]
\]
\begin{equation} \hspace*{4cm}
\label{eqn:B}
 = L_{\pi^+\pi^-}(m|a,p) \cdot \bigg(\frac{2}{3}T^0_0(m) + \frac{1}{3}T^2_0(m)\bigg). 
\end{equation}
Functions $L_{\pi^+\pi^-}(m|a,p)$, $T^0_0(m)$, and $T^2_0(m)$ are described below.
An exotic I=2 S wave $\pi^+_1\pi^+_2\to\pi^+_1\pi^+_2$ scattering is discussed in 
Ref.~\cite{Achasov_PRD67_2003}
\begin{equation}
\label{eqn:E}
E[D^+\to(\pi^+_1\pi^+_2\to\pi^+_1\pi^+_2)\to\pi^+_1\pi^+_2\pi^-, m=m_{\pi^+\pi^+}]
               = L_{\pi^+\pi^+}(m|a,r) \cdot T^2_0(m). 
\end{equation}
It is assumed that the $\sigma$ and $f_0$ mesons can be produced directly 
in the $D^+\to\pi^+\sigma$ and $D^+\to\pi^+f_0$ decays
(we use the ``$DR\pi$'' notation), with an amplitude of
\begin{equation}
\label{eqn:F}
F[D^+\to\pi^+_1(\sigma + f_0)\to\pi^+_1\pi^+_2\pi^-, m=m_{\pi^+_2\pi^-}]
               = T_{0~D^+R\pi^+}^{0~res}(m).
\end{equation}
The point-like $D^+\to\pi^+\pi^0\pi^0$ amplitude is associated with another constant $\bar{a}$
\begin{equation}
\label{eqn:Apl_pi0pi0}
A^{pl}(D^+\to\pi^+\pi^0\pi^0) = 16 \pi \bar{a}.
\end{equation}
Subsequent $\pi^0\pi^0\to\pi^+\pi^-$ rescattering may also contribute to the final state via
the amplitude
\[ \hspace*{-5cm}
\bar{B}[D^+\to\pi^+_1(\pi^0\pi^0\to\pi^+_2\pi^-)\to\pi^+_1\pi^+_2\pi^-, m=m_{\pi^0\pi^0}]
\]
\begin{equation} \hspace*{4cm}
\label{eqn:Bbar}
               = L_{\pi^0\pi^0}(m|\bar{a},q) \cdot \bigg(\frac{2}{3}T^0_0(m) - \frac{2}{3}T^2_0(m)\bigg). 
\end{equation}
In the above equations we assume that $q=r=p$.

The point-like production amplitudes for $D^+\to\pi^+K^+K^-$ and $D^+\to\pi^+K^0\overline{K}^0$
are represented by the constants $c$ and $\bar{c}$,
\begin{equation}
\label{eqn:AplKKpi}
A^{pl}(D^+\to\pi^+K^+K^-) = 16 \pi c.
\end{equation}
\begin{equation}
\label{eqn:AplK0K0pi}
A^{pl}(D^+\to\pi^+K^0\overline{K}^0) = 16 \pi \bar{c}.
\end{equation}
Then, two terms account for the relevant rescattering amplitudes 
$K^+K^-\to\pi^+\pi^-$ and 
$K^0\overline{K}^0\to\pi^+\pi^-$,
\[ \hspace*{-5cm}
C[D^+\to\pi^+_1(K^+K^-\to\pi^+_2\pi^-)\to\pi^+_1\pi^+_2\pi^-,m=m_{K^+K^-}]
\]
\begin{equation} \hspace*{4cm}
\label{eqn:C}
               = L_{K^+K^-}(m|c,s) \cdot T^0_0(K^+K^-\to\pi^+\pi^-,m),
\end{equation}
\[ \hspace*{-5cm}
\bar{C}[D^+\to\pi^+_1(K^0\overline{K}^0\to\pi^+_2\pi^-)\to\pi^+_1\pi^+_2\pi^-, m=m_{K^0\overline{K}^0}]
\]
\begin{equation} \hspace*{4cm}
\label{eqn:Cbar}
                 = L_{K^0\overline{K}^0}(m|\bar{c},t) \cdot T^0_0(K^0\overline{K}^0\to\pi^+\pi^-,m), 
\end{equation}
where we assume that offset parameters are equal, $t=s$.

In above equations we use the function $L_{a\bar{a}}(m|c,d)$, which represents a contribution
from the loop diagram
\begin{equation}
\label{eqn:L}
             L_{a\bar{a}}(m|c,d) =  16 \pi c \cdot \left\{
\begin{array}{ll}
  i\rho_{a\bar{a}}(m) + \rho_{a\bar{a}}(m) \frac{1}{\pi} 
          \ln \frac{1-\rho_{a\bar{a}}(m)}{1+\rho_{a\bar{a}}(m)} + d, 
                                                                              & m \geq 2m_a, \\ 
  -|\rho_{a\bar{a}}(m)| + |\rho_{a\bar{a}}(m)| \frac{2}{\pi} \arctan|\rho_{a\bar{a}}(m)| + d, & m < 2m_a, 
\end{array} \right.
\end{equation}
where  
\begin{equation}
\label{eqn:abs_rho_aabar}
    |\rho_{a\bar{a}}(m)| = \sqrt{4m^2_a/m^2-1},
\end{equation}
\begin{equation}
\label{eqn:rho_aabar}
    \rho_{a\bar{a}}(m) = \sqrt{1-4m^2_a/m^2}.
\end{equation}

Below all definitions, required for parametrization of the amplitude in our case,
are re-written from the recent Ref.~\cite{Achasov_PRD73_2006}.


\subsubsection{$T_0^0 \equiv T_0^0(\pi\pi\to\pi\pi, m)$}

Equation~23 from Ref.~\cite{Achasov_PRD73_2006} gives the S wave amplitude $T_0^0$ of $\pi\pi\to\pi\pi$
scattering with I=0 is
\begin{equation}
\label{eqn:T00}
T_0^0 \equiv T_0^0(\pi\pi\to\pi\pi, m) =
   \frac{\eta_0^0 e^{2i\delta_0^0}-1}{2i\rho_{\pi\pi}(m)}=
   \frac{e^{2i\delta_B^{\pi\pi}}-1}{2i\rho_{\pi\pi}(m)} + 
         e^{2i\delta_B^{\pi\pi}} \cdot T_0^{0~res}(m).
\end{equation}
Equation~24 from Ref.~\cite{Achasov_PRD73_2006} gives the total phase
\begin{equation}
\label{eqn:delta00}
    \delta_0^0 = \delta_0^0(m) = \delta_B^{\pi\pi}(m) + \delta_{res}(m).
\end{equation}
Equation~25 from Ref.~\cite{Achasov_PRD73_2006} defines the resonant part of the
$S$ matrix
\begin{equation}
\label{eqn:S00_res}
S_0^{0~res}(m) = \eta_0^0(m) e^{2i\delta_{res}(m)}
               = 1 + 2i\rho_{\pi\pi}(m) \cdot T_0^{0~res}(m),
\end{equation}
which can be described by the inelasticity
\begin{equation}
\label{eqn:eta00}  \eta_0^0(m) = |S_0^{0~res}(m)|
\end{equation}
and resonant phase
\begin{equation}
\label{eqn:delta_res}
    \delta_{res}(m) = \frac{1}{2}\cdot \arctan\bigg( \frac{\Im(S_0^{0~res})}{\Re(S_0^{0~res})} \bigg).    
\end{equation}

The chiral background shielding phase
$\delta_B^{\pi\pi}(m)$, motivated by the $\sigma$ model, is taken as
Eq.~26 from Ref.~\cite{Achasov_PRD73_2006}:
\begin{equation}
\label{eqn:tan_delta_B}
    \tan\delta_B^{\pi\pi} = -\frac{p_\pi}{m_\pi} 
                            \bigg( b_0 - b_1 \frac{p^2_\pi}{m^2_\pi} + b_2 \frac{p^4_\pi}{m^4_\pi}     
                            \bigg) \times
                            \frac{1}{1 + (2 p_\pi)^2/\Lambda^2},
\end{equation}
where $2 p_\pi = \sqrt{m^2-4m_\pi^2}$, and $(1 + (2 p_\pi)^2/\Lambda^2)^{-1}$ is a cutoff factor.
The value of parameters $b_0$, $b_1$, $b_2$, and $\Lambda$ used in our fits are listed in 
Table~\ref{tab:A_model_pars}.
The background phase is derived from Eq.~\ref{eqn:tan_delta_B}
\begin{equation}
\label{eqn:delta_B}
    \delta_B^{\pi\pi} = \arctan [\tan\delta_B^{\pi\pi}].
\end{equation}

\subsubsection{Resonance amplitudes $T_{0~D^+R\pi^+}^{0~res}(m)$ and $T_0^{0~res}(m)$}

In Eq.~\ref{eqn:F}, \ref{eqn:T00}, and~\ref{eqn:S00_res}
we use a brief notation for the production and scattering resonance amplitudes 
expressed through the mixing matrix operator $G^{-1}_{RR'}$
\begin{equation}
\label{eqn:T00_res_DRpi}
T_{0~D^+R\pi^+}^{0~res} =  e^{i\delta_B^{\pi\pi}(m)} 
                           \sum_{RR'} \frac{g_{D^+R\pi^+} G^{-1}_{RR'} g_{R'\pi^+\pi^-}}{16\pi},
\end{equation}
\begin{equation}
\label{eqn:T00_res}
T_0^{0~res} = \sum_{RR'} \frac{g_{R\pi\pi} G^{-1}_{RR'} g_{R'\pi\pi}}{16\pi}.
\end{equation}
Note the difference between specific coupling constants 
and the exponential factor in 
Eqs.~\ref{eqn:T00_res_DRpi} and \ref{eqn:T00_res}.

\subsubsection{$T_0^0(K\overline{K}\to\pi\pi, m)$}

The S wave amplitude of $K\overline{K}\to\pi\pi$ scattering, taking in to account mixing
through $RR'$ resonances (i.e. $\sigma$ and $f_0(980)$ mesons) is given by
Eq.~3 from Ref.~\cite{Achasov_PRD73_2006}: 
\begin{equation}
\label{eqn:T00_KcKc-picpic}
  T_0^0(K^+K^-\to\pi^+\pi^-, m) =
         e^{i\delta_B} \sum_{RR'} \frac{g_{RK^+K^-} G^{-1}_{RR'} g_{R'\pi^+\pi^-}}{16\pi},
\end{equation}
\begin{equation}
\label{eqn:T00_K0K0-picpic}
  T_0^0(K^0\overline{K}^0\to\pi^+\pi^-, m) =
         e^{i\delta_B} \sum_{RR'} \frac{g_{RK^0\overline{K}^0} G^{-1}_{RR'} g_{R'\pi^+\pi^-}}{16\pi},
\end{equation}
where Eq.~4 from Ref.~\cite{Achasov_PRD73_2006} defines 
\begin{equation}
\label{eqn:delta_B_tot}
  \delta_B = \delta_B^{\pi\pi} + \delta_B^{K\overline{K}}.
\end{equation}
Equation~28 from Ref.~\cite{Achasov_PRD73_2006} is
\begin{equation}
\label{eqn:tan_delta_BKK-pipi}
  \tan \delta_B^{K\overline{K}} = f_K(m)\cdot 2p_K = f_K(m)\cdot\sqrt{m^2-4m^2_K},
\end{equation}
where Eq.~36 from Ref.~\cite{Achasov_PRD73_2006} gives
\begin{equation}
\label{eqn:f_K}
  f_K(m) = -\arctan\bigg(\frac{m^2-m^2_1}{m^2_2} \bigg)\bigg/\Lambda_K,
\end{equation}
and we find the phase as
\begin{equation}
\label{eqn:delta_BKK-pipi}
  \delta_B^{K\overline{K}} = \arctan ( \tan \delta_B^{K\overline{K}} ).
\end{equation}
The value of parameters  $m_1$, $m_2$, and $\Lambda_K$ used in our fits are listed in 
Table~\ref{tab:A_model_pars}.

\subsubsection{An exotic I=2 amplitude $T_0^2(m)\equiv T_0^2(\pi^+\pi^+ \to \pi^+\pi^+, m)$}
According to Ref.~\cite{Achasov_PRD67_2003} the  I=2 $\pi^+\pi^+ \to \pi^+\pi^+$
the rescattering amplitude is given in a unitarian form
\begin{equation}
   \label{eqn:T02}
  T_0^2(m)\equiv T_0^2(\pi^+\pi^+ \to \pi^+\pi^+, m) =
                   \frac{\eta_0^2(m) e^{2i\delta_0^2(m)}-1}{2i}.
\end{equation}
The phase shift $\delta_0^2(m)$ is parameterized by
\begin{equation}
   \label{eqn:delta_I2}
   \delta_0^2(m) = \frac{-a\sqrt{m^2/4-m_{\pi}^2}}{1 + bm^2 + cm^4 + dm^6}.
\end{equation}
From fit in Ref.~\cite{Achasov_PRD67_2003}
to data for the $\pi^-p \to \pi^0\pi^0 n$ process in Refs~\cite{Hoogland_NPB126_1977}
and \cite{Durusoy_PL45B_1973},
the parameters of Eq.~\ref{eqn:delta_I2} are
  $a = (55.21 \pm 3.18)$~deg/GeV,       
  $b = (0.853 \pm 0.254)$~GeV$^{-2}$,   
  $c = (-0.959 \pm 0.247)$~GeV$^{-4}$, and  
  $d = (0.314 \pm 0.070)$~GeV$^{-6}$.

The $\eta_0^2(m)$ is an inelasticity for the wave with total spin $0$ and
isospin $2$. In the mass range of $m<m(\rho\rho)$ ($\sim$1.54~GeV)
the inelasticity parameter $\eta_0^2(m)$
should be represented by the smooth real function of $m$.
An appropriate fit to data has been considered in Ref.~\cite{Zou_Bugg_2004},
see their Fig.~2, and we use the approximation
\begin{equation}
\label{eqn:eta02}
             \eta_0^2(m) = \left\{
\begin{array}{ll}
                                1, & m \leq 1~{\rm GeV}/c^2 \\ 
                                \propto \cos {\rm -like~~smooth~~transition}, & 1<m<1.7~{\rm GeV/c}^2 \\ 
                                0.4, & m \geq 1.7~{\rm GeV}/c^2. \\ 
\end{array} \right.
\end{equation}
In our case
we neglect the small $D$ wave scattering amplitude 
$T_2^2(\pi^+\pi^+ \to \pi^+\pi^+)$.

\subsubsection{Mixing matrix $G_{RR'}(m)$}

The mixing operator $G_{RR'}(m)$ is a matrix of inverse propagators,
with rank equal to the number of mixed resonances.
In case of mixing of two resonances $R$ and $R'$ this matrix has the form,
following Eq.~5 of Ref.~\cite{Achasov_PRD73_2006},
\begin{equation}
\label{eqn:G-matrix}
  G_{RR'}(m) = \left(
               \begin{array}{cc}
                      D_{R}(m)      & -\Pi_{RR'}(m) \\
                 -\Pi_{R'R}(m)      &    D_{R'}(m)
               \end{array}
              \right).
\end{equation}
In general, the diagonal elements of this matrix are the
inverse propagators
\begin{equation}
  D_R(m) = m_R^2 - m^2 -im\Gamma_R(m),
\end{equation}
while the non-diagonal elements are polarization operators
describing mixing.
An expression for the inverse propagator of the scalar resonance is given in
Eq.~6 from Ref.~\cite{Achasov_PRD73_2006},
\begin{equation}
  \label{eqn:scal_propagator}
     D_{R}(m) = m_R^2 - m^2 + \sum_{ab} g_{Rab} [Re P_R^{ab}(m_R) - P_R^{ab}(m)],
\end{equation}
where $\sum_{ab} g_{Rab} [Re P_R^{ab}(m_R) - P_R^{ab}(m)] = Re[\Pi_R(m_R)] - \Pi_R(m)$
takes in to account the finite width correction.
After Eq.~5 in Ref.~\cite{Achasov_PRD73_2006} the non-diagonal terms of the
polarization operator are given by equation
\begin{equation}
  \label{eqn:pi_RRprime}
     \Pi_{RR'}(m) = \sum_{ab} g_{R'ab} P_R^{ab}(m) + C_{RR'},
\end{equation}
where the constants $C_{RR'}$ take into account effectively the contribution
of $VV$, 4$P$ and other intermediate states and incorporates the subtraction constants for the
$R \to (PP) \to R'$ transitions.
Here we use the notation from different publications, \cite{Achasov_PRD55_1997}--\cite{Achasov_PRD73_2006},
\begin{equation}
     P_R^{ab}(m)=
          \frac{g_{Rab}}{16\pi^2} P^{ab}(m),  ~~{\rm or} ~~
     \Pi_R^{ab}(m)=
          \frac{g^2_{Rab}}{16\pi^2} P^{ab}(m),
\end{equation}
and
\begin{equation} 
     \Pi_R(m)= \sum_{ab}\Pi_R^{ab}(m).
\end{equation}
Eqs.~7--9 from Ref.~\cite{Achasov_PRD73_2006}
(also Ref.~\cite{Achasov_PRD56_1997}, Eq.~30 
and Ref.~\cite{Achasov_PRD70_2004}, Eqs.~16,19,22) for
$m_a>m_b$,
$m_+=m_a+m_b$, and
$m_-=m_a-m_b$
give
\begin{equation}
     P^{ab}(m)= \frac{m_+m_-}{m^2} \ln \frac{m_b}{m_a} +
               \left\{
               \begin{array}{ll}
                    \rho_{ab}(m)\cdot \bigg[ i\pi +
                                       \ln\frac{\sqrt{m^2-m_-^2}-\sqrt{m^2-m_+^2}}
                                               {\sqrt{m^2-m_-^2}+\sqrt{m^2-m_+^2}} \bigg],
                    & m > m_+\\
                    - \pi |\rho_{ab}(m)| + 2|\rho_{ab}(m)|\arctan\frac{\sqrt{m_+^2-m^2}}
                                                {\sqrt{m^2-m_-^2}},
                    & m_- \leq m \leq m_+\\
                    - \rho_{ab}(m)\cdot \ln\frac{\sqrt{m_+^2-m^2}-\sqrt{m_-^2-m^2}}
                                                {\sqrt{m_+^2-m^2}+\sqrt{m_-^2-m^2}},
                    & m < m_-\\
            \end{array}
              \right.
\end{equation}

\begin{equation}
     \rho_{ab}(m) = \sqrt{\bigg(1-\frac{m_+^2}{m^2}\bigg)\bigg(1-\frac{m_-^2}{m^2}\bigg)}.
\end{equation}

The constants $g_{Rab}$ are related to the width, Eq.~11 from Ref.~\cite{Achasov_PRD73_2006},
\begin{equation}
\label{eqn:Gamma_and_g_Rab}
     \Gamma(R\to ab, m) = \frac{g^2_{Rab}}{16\pi m} \rho_{ab}(m).
\end{equation}

\subsubsection{Model parameters}

In the mixing operator Eq.~\ref{eqn:G-matrix} we account for 
seven intermediate states:
$\pi^+\pi^-$, 
$\pi^0\pi^0$, 
$K^+K^-$, 
$K^0\overline{K}^0$, 
$\eta\eta$, 
$\eta'\eta$, and
$\eta'\eta'$.
We follow the conventions of Ref.~\cite{Achasov_PRD73_2006}
for coupling constants, motivated by the four-quark model.
For the $f_0(980)$ and similarly for the $\sigma$ we use
\begin{equation}
\label{eqn:g_f0pipi_KK}
g_{f_0K^0\bar{K^0}} = g_{f_0K^+K^-}, ~~~
g_{R\pi^0\pi^0} = g_{R\pi^+\pi^-}  / \sqrt{2}, ~~~
g_{R\pi\pi} = \sqrt{3/2}g_{R\pi^+\pi^-}.
\end{equation}

For the $f_0(980)$ coupling constants to $\eta^{(\prime)}\eta^{(\prime)}$ we use
\begin{equation}
\label{eqn:g_f0etaeta}
g_{f_0\eta\eta} = -g_{f_0\eta'\eta'} = \frac{2\sqrt{2}}{3} g_{f_0K^+K^-}, ~~~
g_{f_0\eta'\eta} = - \frac{\sqrt{2}}{3} g_{f_0K^+K^-}.
\end{equation}

For the $\sigma$ coupling constants to $\eta^{(\prime)}\eta^{(\prime)}$ we use
\begin{equation}
\label{eqn:g_sigmaetaeta}
g_{\sigma\eta\eta} = g_{\sigma\eta'\eta'} = \frac{\sqrt{2}}{3} g_{\sigma\pi^+\pi^-}, ~~~
g_{\sigma\eta'\eta} = \frac{1}{3\sqrt{2}} g_{\sigma\pi^+\pi^-}.
\end{equation}

Further
we use the values of the parameters shown in Table~\ref{tab:A_model_pars},
which are taken from Fit~1 of Ref.~\cite{Achasov_PRD73_2006}.
\begin{table}[!htb]
\caption{\label{tab:A_model_pars} Achasov model parameters from Fit~1 of Ref.~\cite{Achasov_PRD73_2006} 
                                  used in our calculations.}
\begin{center}
\begin{tabular}{|c|c||c|c|}
\hline
\hline
Parameter                    & Value in Fit~1 \cite{Achasov_PRD73_2006}  
                             & Parameter                    
                             & Value in Fit~1 \cite{Achasov_PRD73_2006} \\
\hline
$m_{f_0}$, MeV               & 984.1  & $b_0$                        & 4.9   \\
$m_{\sigma}$, MeV            & 461.9  & $b_1$                        & 1.1   \\
$g_{f_0K^+K^-}$, GeV         & 4.3    & $b_2$                        & 1.36  \\
$g_{f_0\pi^+\pi^-}$, GeV     &--1.8   & $\Lambda$, MeV               & 172.2 \\
$g_{\sigma K^+K^-}$, GeV     & 0.55   & $m_1$, MeV                   & 765.4 \\
$g_{\sigma \pi^+\pi^-}$, GeV & 2.4    & $m_2$, MeV                   & 368.9 \\
$C_{f_0\sigma}$              &--0.047 \cite{Achasov_typo}
                                      & $\Lambda_K$, GeV             & 1.24  \\
\hline
\hline
\end{tabular}
\end{center}
\end{table}


\begin{figure}[!htb]
 \begin{minipage}[t]{75mm}
  \includegraphics[width=70mm]{\FigsAch/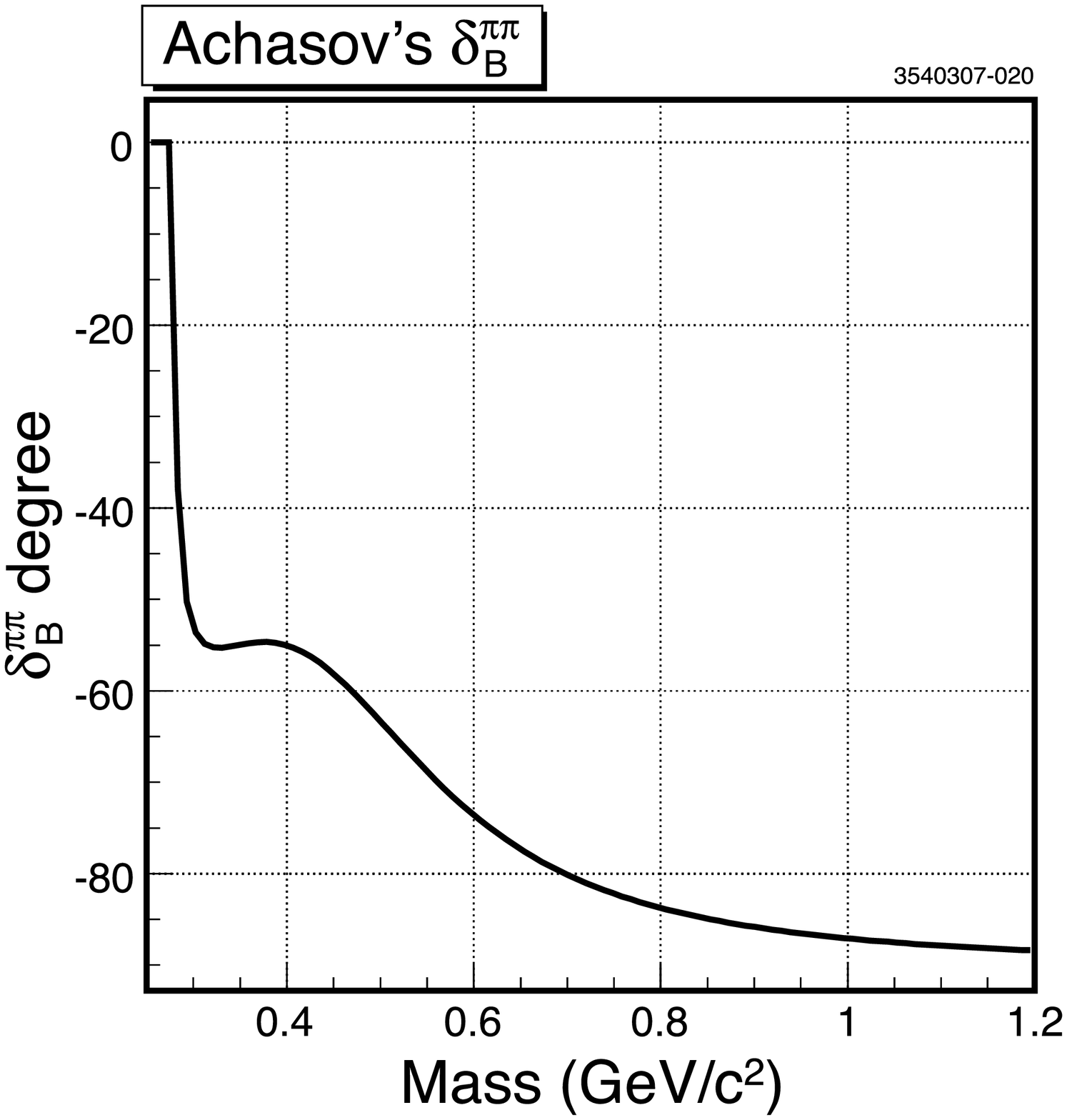}
  \caption{\label{fig:delta_B_pipi} 
           The background phase in $\pi\pi$ scattering, $\delta_B^{\pi\pi}(m)$, 
           from Eq.~\ref{eqn:delta_B}.}
 \end{minipage}
 \hfill
 \begin{minipage}[t]{75mm}
  \includegraphics[width=70mm]{\FigsAch/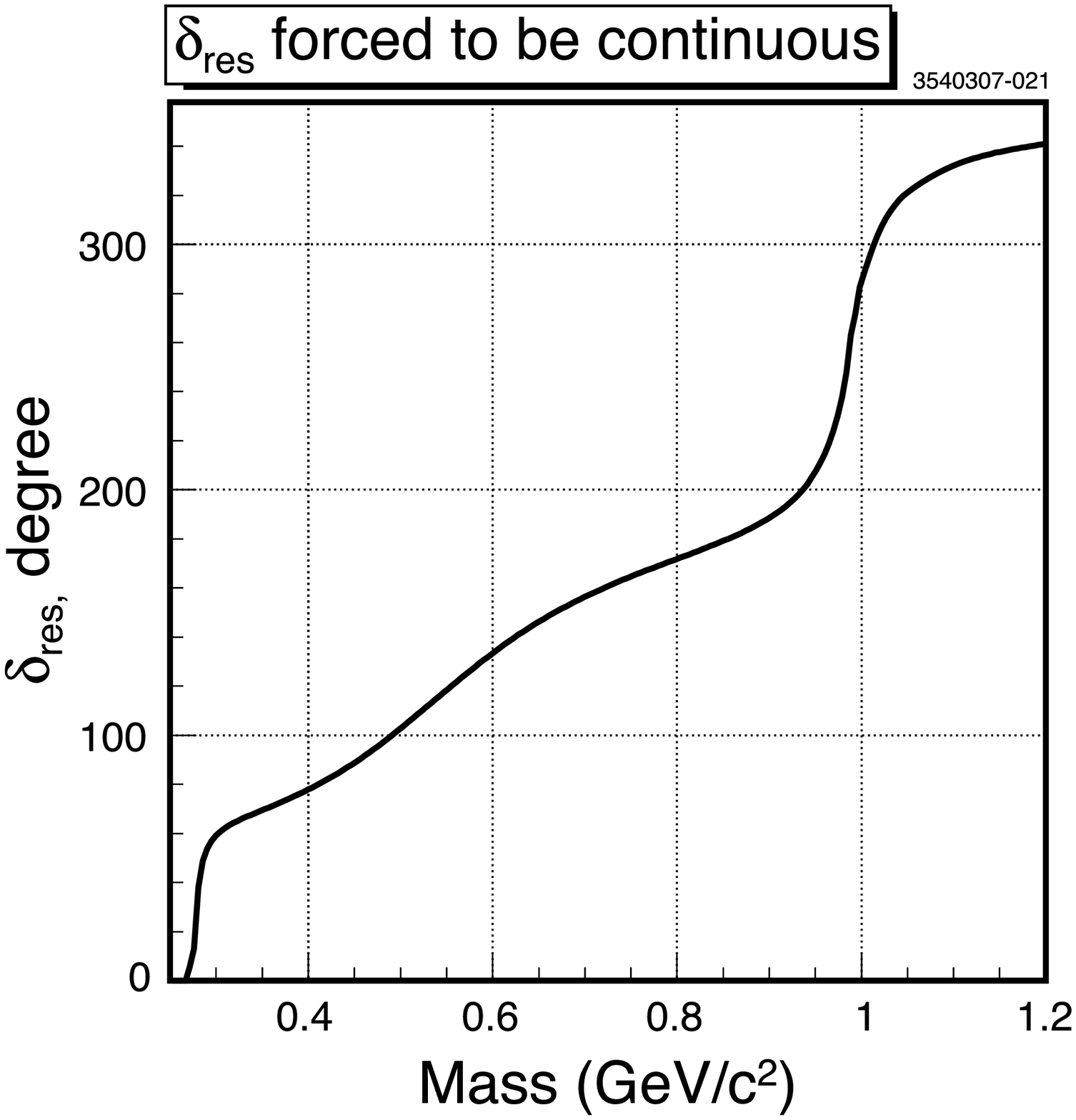}
  \caption{\label{fig:delta_res_corrected} 
           The phase of the resonance $\pi\pi$ scattering, 
           $\delta_{res}(m)$, from Eq.~\ref{eqn:delta_res}.
          }
 \end{minipage}
 \begin{minipage}[t]{75mm}
  \includegraphics[width=70mm]{\FigsAch/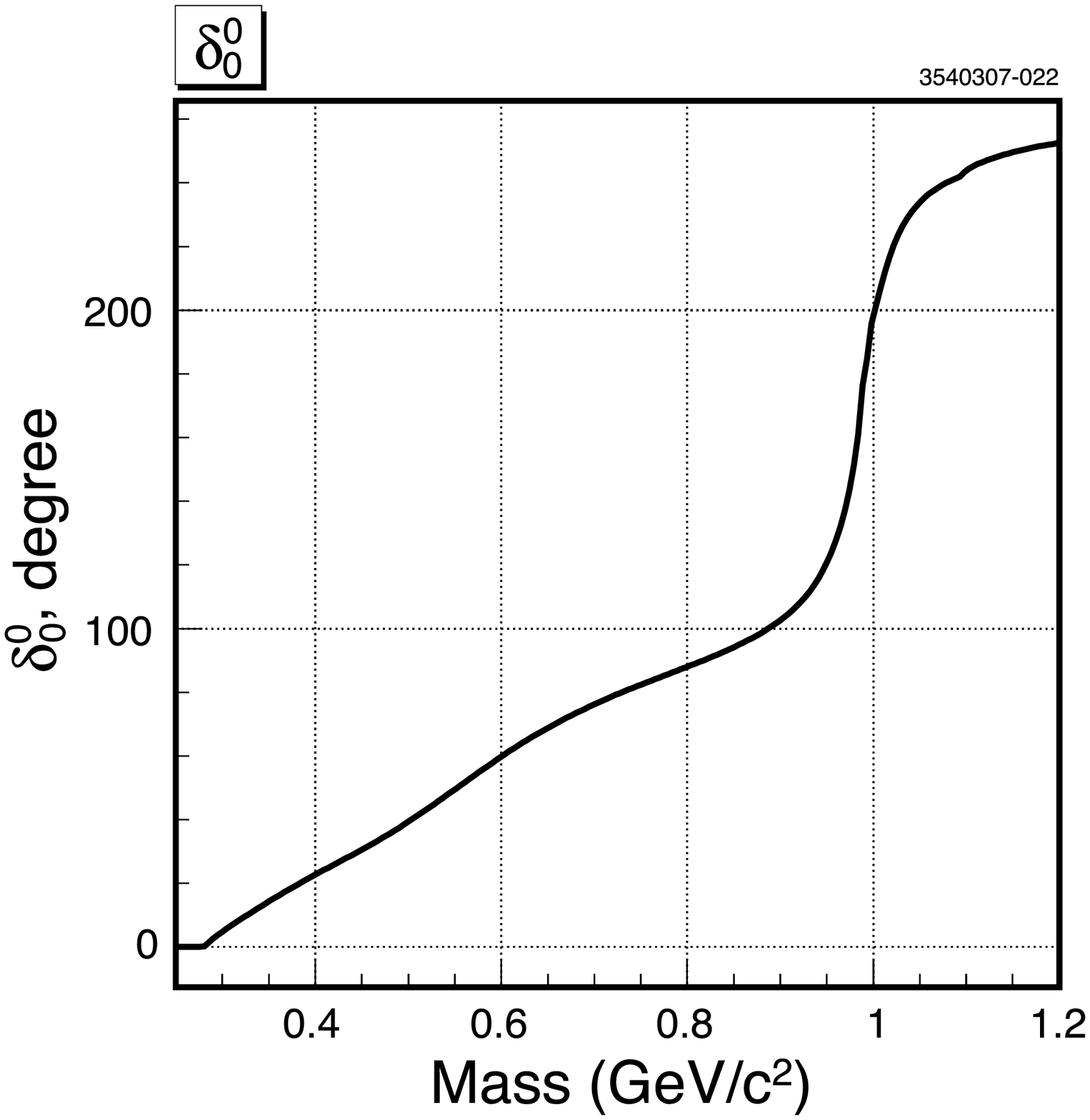}
  \caption{\label{fig:delta00} 
           The total $\pi\pi$ scattering phase, 
           $\delta_0^0(m)$, from Eq.~\ref{eqn:delta00}.
          }
 \end{minipage}
 \hfill
 \begin{minipage}[t]{75mm}
  \includegraphics[width=70mm]{\FigsAch/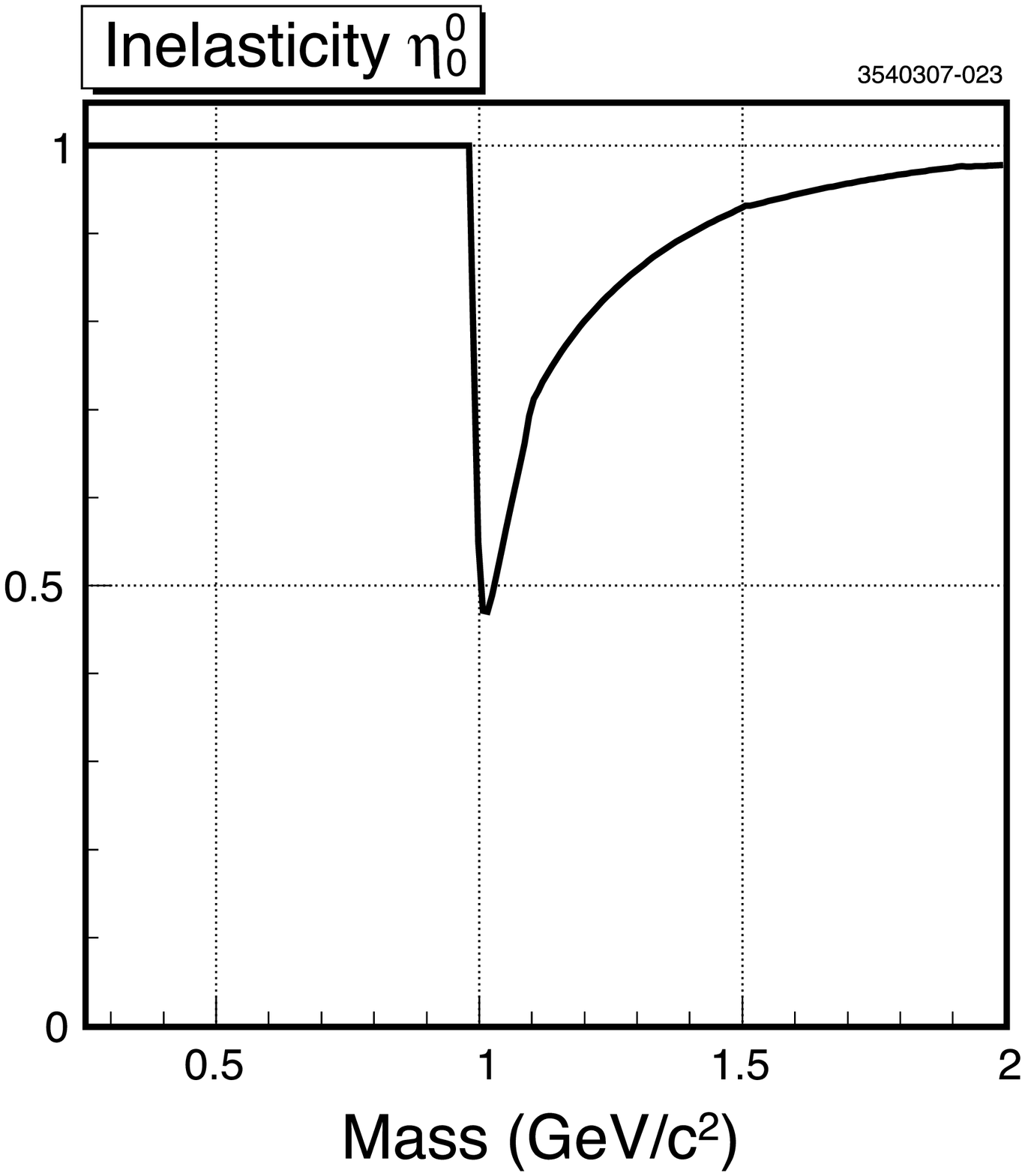}
  \caption{\label{fig:eta00} 
           The inelasticity,
           $\eta_0^0$, from Eq.~\ref{eqn:eta00} for $2m_\pi<m<2$~GeV/$c^2$.
          }
 \end{minipage}
\end{figure}


\begin{figure}[!htb]
 \begin{minipage}[t]{75mm}
  \includegraphics[width=70mm]{\FigsAch/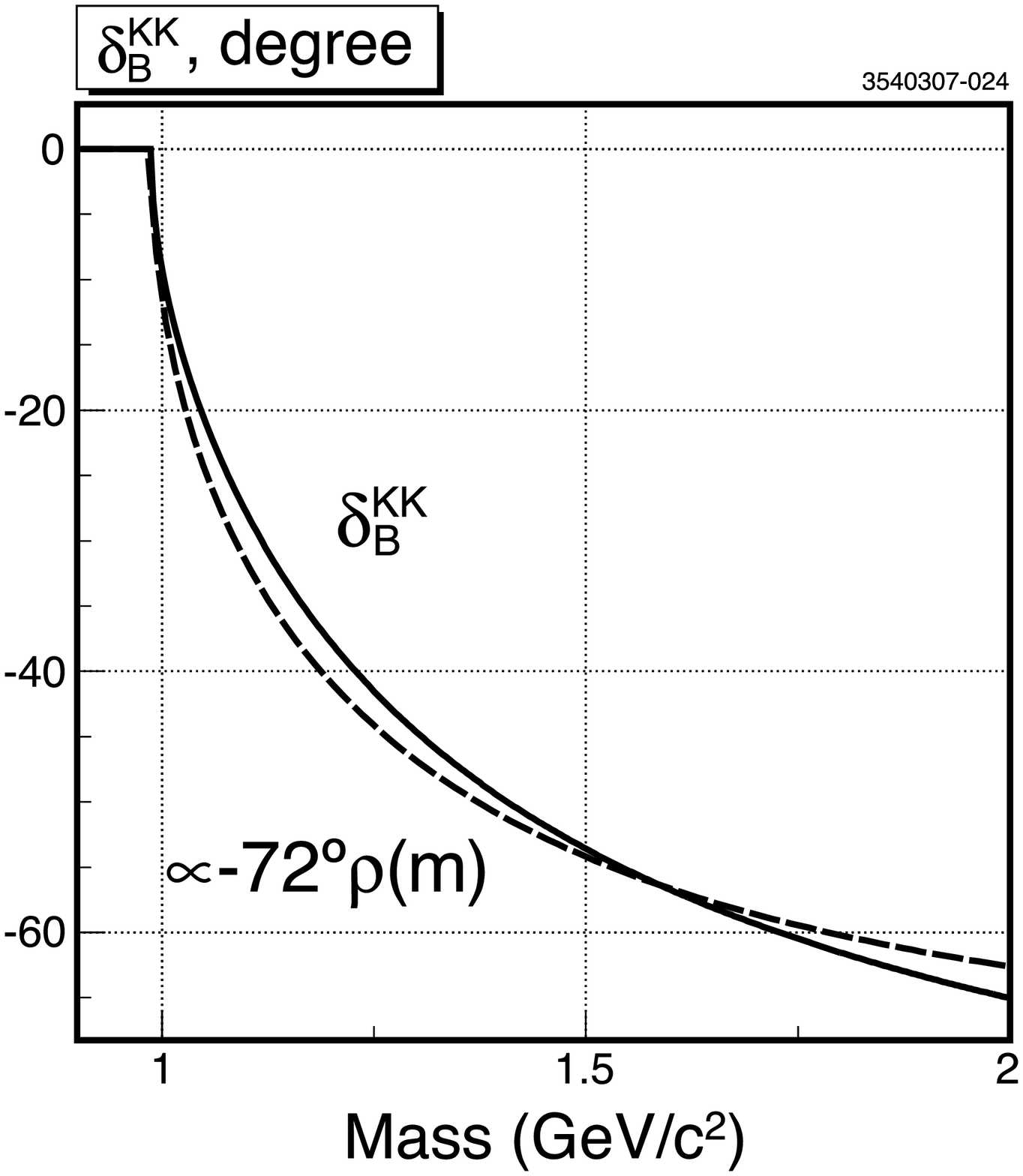}
  \caption{\label{fig:delta_B_KK} 
           The background phase in $K\bar{K}$ scattering,
           $\delta_B^{K\bar{K}}$, from Eq.~\ref{eqn:delta_BKK-pipi} (solid curve),
           and its approximation by the phase space factor (dashed curve).
          }
 \end{minipage}
 \hfill
 \begin{minipage}[t]{75mm}
  \includegraphics[width=70mm]{\FigsAch/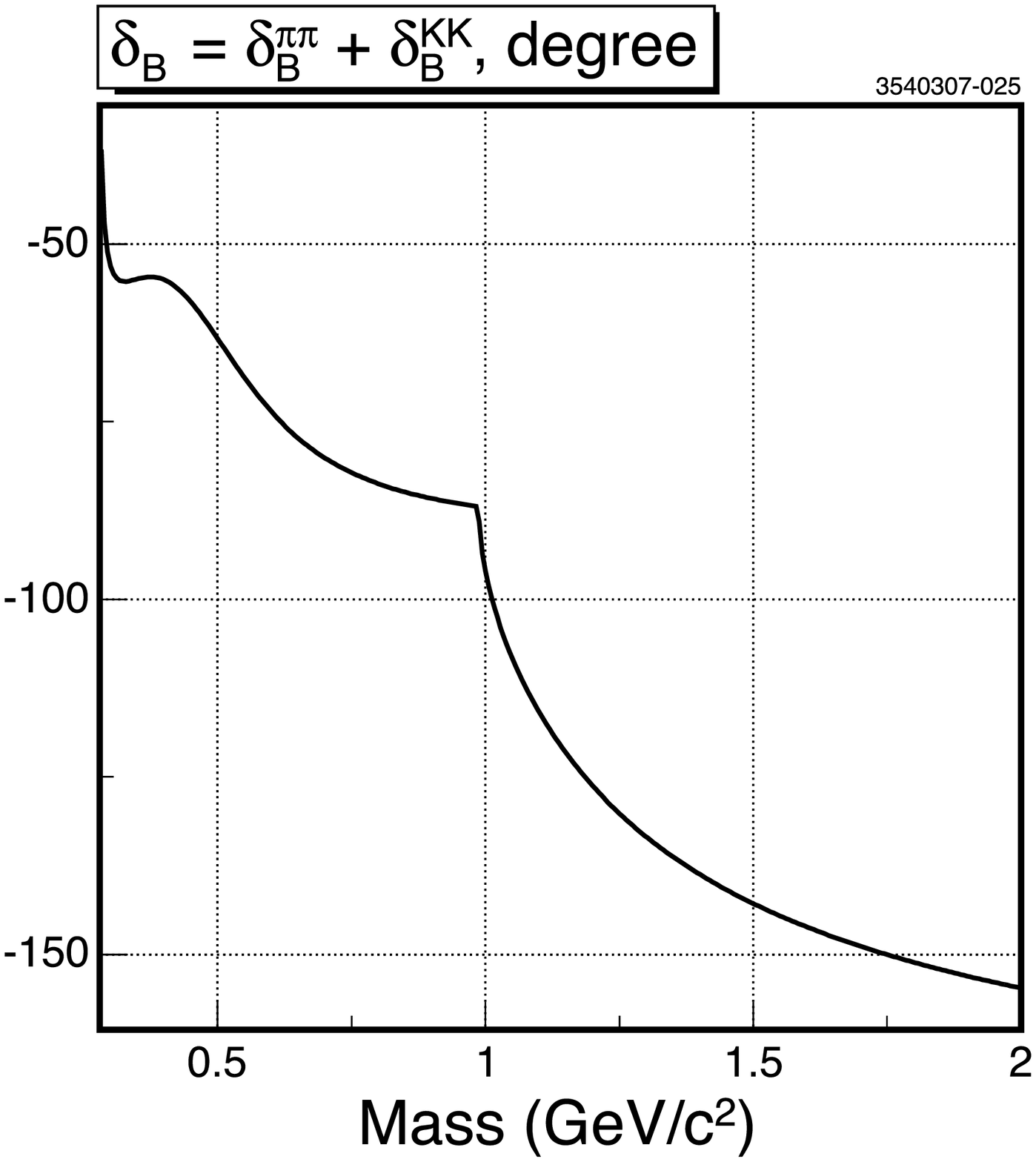}
  \caption{\label{fig:delta_B_pipi_KK} 
           The total background phase in $K\bar{K} \to \pi\pi$ scattering, 
           $\delta_B = \delta_B^{\pi\pi} + \delta_B^{K\bar{K}}$,
           from Eq.~\ref{eqn:delta_B_tot}.}
 \end{minipage}
\end{figure}
\begin{figure}[!htb]
 \begin{minipage}[t]{75mm}
  \includegraphics[width=70mm]{\FigsAch/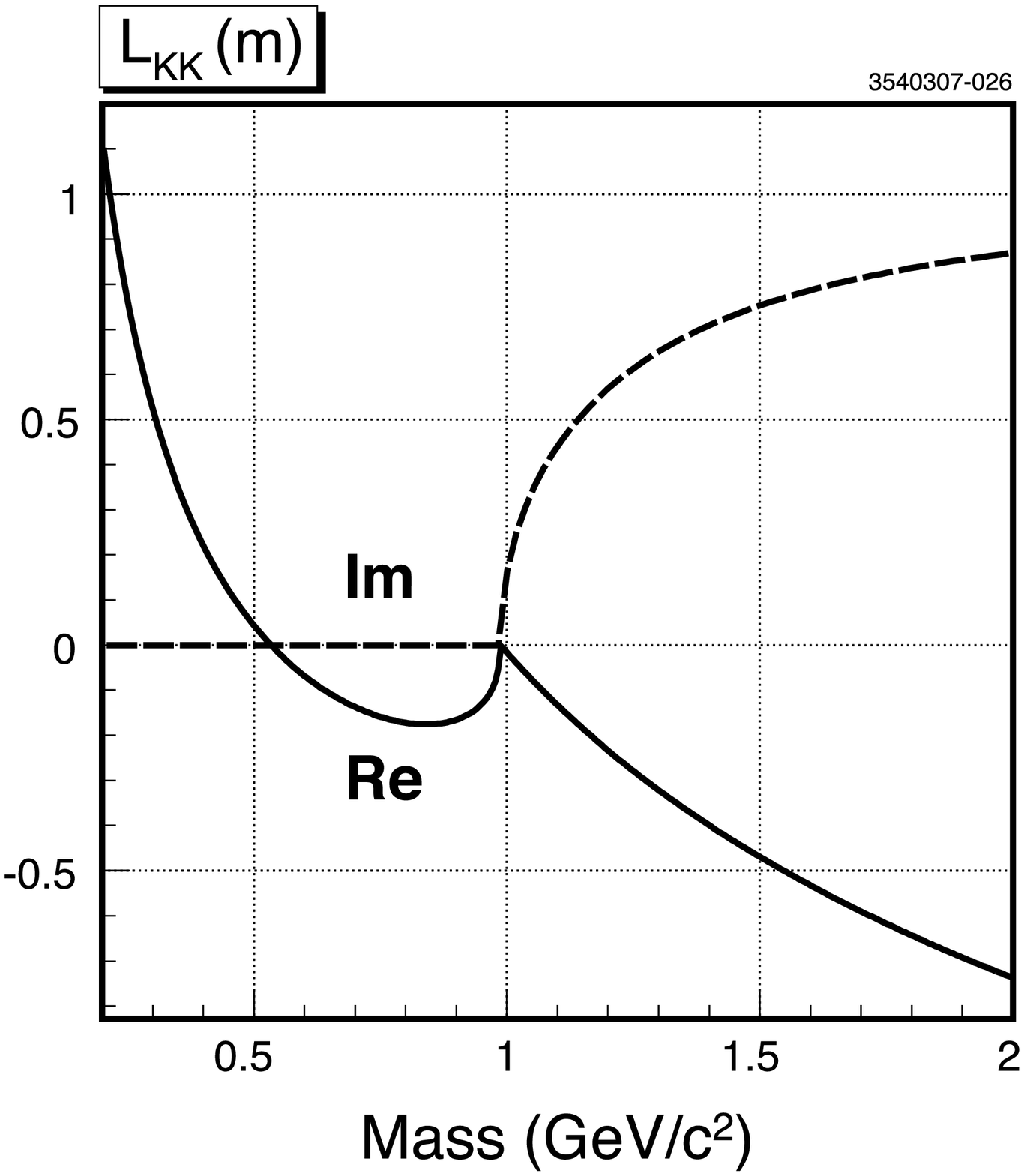}
  \caption{\label{fig:L_KcKc} 
           The loop integral, $L_{K^+K^-}(m,1,0) / 16\pi$, from Eq.~\ref{eqn:L}.
	   The real (solid curve) and imaginary (dashed curve) parts of the complex function are shown.
          } 
 \end{minipage}
 \hfill
 \begin{minipage}[t]{75mm}
  \includegraphics[width=70mm]{\FigsAch/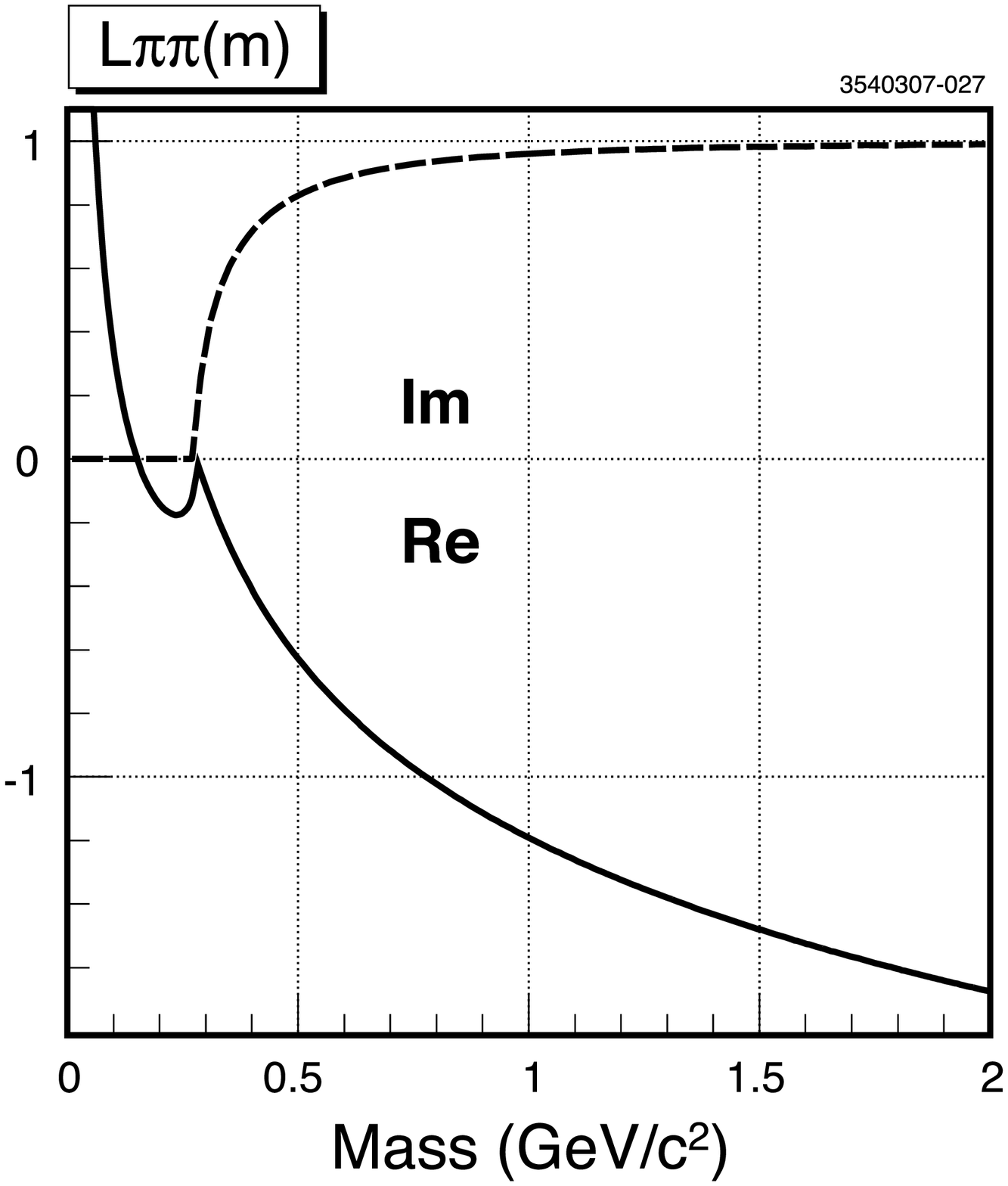}
  \caption{\label{fig:L_picpic} 
           The loop integral, $L_{\pi^+\pi^-}(m,1,0) / 16\pi$, from Eq.~\ref{eqn:L}.
           The real (solid curve) and imaginary (dashed curve) parts of the complex function are shown. 
          }
 \end{minipage}
\end{figure}

\subsubsection{Check for $\delta_{res}(m)$, $\delta_B^{\pi\pi}(m)$, $\delta_0^0(m)$, $\eta_0^0(m)$, etc. }
\label{sec:complex_amplitudes}
In order to check that the code for this parameterization works 
properly we reproduce plots from Ref.~\cite{Achasov_PRD73_2006}. 
\begin{itemize}
\item
[$\mathbf{\delta_{res}(m)}$:]
We define the $\delta_{res}(m)$ as the phase of the
complex function $S_0^{0~res}(m)$ in Eq.~\ref{eqn:delta_res}.
However, this phase has discontinuities
in the vicinity of each resonance mass, but not exactly at the resonance mass value.
In further calculations we require that the phase is continuous, as shown in
Fig.~\ref{fig:delta_res_corrected},
by adding a phase shift of $\pi$ above each discontinuity point.
This plot is consistent with Fig.~3 in Ref.~\cite{Achasov_PRD73_2006}.
\item
[$\mathbf{\delta_B^{\pi\pi}(m)}$:]
The background phase $\delta_B^{\pi\pi}(m)$ is derived from
Eq.~\ref{eqn:delta_B}, as shown in Fig.~\ref{fig:delta_B_pipi}.
This plot is consistent with Fig.~2 in Ref.~\cite{Achasov_PRD73_2006}.
\item
[$\mathbf{\delta_0^0(m)}$:]
The total phase $\delta_0^0(m)$ represented by Eq.~\ref{eqn:delta00}
is shown in Fig.~\ref{fig:delta00}. This plot is consistent with
Fig.~4 in Ref.~\cite{Achasov_PRD73_2006}.
\item
[$\mathbf{\eta_0^0(m)}$:]
The $\eta_0^0(m)$ derived from Eq.~\ref{eqn:eta00} is displayed in 
Fig.~\ref{fig:eta00} which shows that $\eta_0^0(m)=1$ at $m<m_{K\bar K}$
confirming unitarity in $\pi\pi \to \pi\pi$ scattering,
consistent with Fig.~6 from Ref.~\cite{Achasov_PRD73_2006}.
\end{itemize}


We also tested all complex functions and their components from 
Eq.~\ref{eqn:amp_D3pi}. In particular,
Fig.~\ref{fig:delta_B_KK} shows $\delta_B^{K\bar{K}}$ from Eq.~\ref{eqn:delta_BKK-pipi};
Fig.~\ref{fig:delta_B_pipi_KK} shows  $\delta_B$ from Eq.~\ref{eqn:delta_B_tot};
Figs.~\ref{fig:L_KcKc}, \ref{fig:L_picpic} show the loop integrals 
$L_{K^+K^-}(m|1,0) / 16\pi$ and  $L_{\pi^+\pi^-}(m|1,0) / 16\pi$, respectively, 
from Eq.~\ref{eqn:L}.


\subsubsection{S wave implementation in the code of the Dalitz plot fitter}

As usually in a Dalitz plot analysis each amplitude fraction is taken with its own complex coefficient 
$c_{\rm{mode}} = a_{\rm{mode}} e^{i \phi_{\rm{mode}}}$
represented by two real numbers, an amplitude $a_{\rm{mode}}$ and phase $\phi_{\rm{mode}}$.
The loop integral in Eq.~\ref{eqn:L} has an additional offset constant
$d_{\rm{mode}}$. Unitarity requires that $d_{\rm{mode}}$ is real.
All these constants, as well as unknown coupling constants
$g_{D^+\sigma\pi^+}$ and $g_{D^+f_0\pi^+}$ from 
Eq.~\ref{eqn:T00_res_DRpi},
are the fit parameters, which can be free to float or fixed.
The actual parameterization for 
$m=m_x\equiv \min[m(\pi^+_1\pi^-), m(\pi^+_2\pi^-)]$ or 
$m=m_y\equiv \max[m(\pi^+_1\pi^-), m(\pi^+_2\pi^-)]$ is given by the amplitude 
\begin{eqnarray}
   \label{eqn:amp_picpic}
    A_{\pi^+\pi^-}(m)
        & = & 16 \pi c_{\pi\pi} 
              \label{eqn:amp_pipi_const} \\        
        & + & L_{\pi^+\pi^-}(m | c_{\pi\pi}, d_{\pi\pi})
              \cdot \bigg(\frac{2}{3}T_0^0(m) + \frac{1}{3}T_0^2(m) \bigg)
              \label{eqn:amp_picpic_picpic} \\
        & + & L_{\pi^0\pi^0}(m | c_{\pi^0\pi^0}, d_{\pi^0\pi^0})
              \cdot \bigg(\frac{2}{3}T_0^0(m) - \frac{2}{3}T_0^2(m) \bigg)
              \label{eqn:amp_pi0pi0_picpic} \\
        & + & L_{K^+K^-}(m | c_{K^+K^-}, d_{K^+K^-}) 
              \cdot T_0^0(K^+K^- \to \pi^+\pi^-, m)
              \label{eqn:amp_KcKc_picpic} \\
        & + & L_{K^0\overline{K}^0}(m | c_{K^0\overline{K}^0}, d_{K^0\overline{K}^0}) 
              \cdot T_0^0(K^0\overline{K}^0 \to \pi^+\pi^-, m)
              \label{eqn:amp_K0K0_picpic} \\
        & + & c_{D^+R\pi^+} 
              \cdot T_{0~DR\pi}^{0~res}(m)
              \label{eqn:amp_DRpi_picpic}.   
\end{eqnarray}
The I=2 $\pi^+\pi^+ \to \pi^+\pi^+$ scattering amplitude for $m=m_z\equiv m(\pi^+_1\pi^+_2)$ 
is given by 
\begin{equation}
\label{eqn:amp_picpic_I2}
    A_{\pi^+\pi^+}(m) = L_{\pi^+\pi^+}(m | c_{\pi\pi}, d_{\pi\pi})\cdot T_0^2(m).
\end{equation}
It is worth noting that three terms in
Eqs.~\ref{eqn:amp_pipi_const},
    \ref{eqn:amp_picpic_picpic} and 
    \ref{eqn:amp_picpic_I2} 
have a common complex coefficient $c_{\pi\pi}$,
appearing from the point-like term, and two of them have a common offset parameter $d_{\pi\pi}$
from the loop integral.
The total contribution of Achasov's S wave in the Dalitz plot amplitude is
\begin{equation}
\label{eqn:Swave_amp_total}
    A_{SW}(m_x,m_y,m_z) = A_{\pi^+\pi^-}(m_x) 
                        + A_{\pi^+\pi^-}(m_y)
                        + A_{\pi^+\pi^+}(m_z).
\end{equation}
The ``$DR\pi$'' sub-mode in Eq.~\ref{eqn:amp_DRpi_picpic}
has a redundant freedom for amplitude factors due to the products
$a_{D^+R\pi^+} \cdot g_{D^+\sigma\pi^+}$ and $a_{D^+R\pi^+} \cdot g_{D^+f_0\pi^+}$.
In our fits we fix $a_{D^+R\pi^+}=1$, or $a_{D^+R\pi^+}=0$ to turn it off,
and use coupling constants $g_{D^+\sigma\pi^+}$ and $g_{D^+f_0\pi^+}$.

For a first approximation we try to eliminate the number of free parameters in the function.
We assume $d_{\pi^0\pi^0} = d_{\pi\pi}$ and $d_{K^0\overline{K}^0} = d_{K^+K^-}$
from isospin symmetry.
We note that the parameterization for 
$K^0\overline{K}^0 \to \pi^+\pi^-$ in Eq.~\ref{eqn:amp_K0K0_picpic}
is nearly the same as that for 
$K^+K^- \to \pi^+\pi^-$ in Eq.~\ref{eqn:amp_KcKc_picpic}.
The small difference appears due to the different masses of the $K^+$ and $K^0$ mesons. 
Keeping in mind this small difference between amplitudes we do not consider separate contributions
from $K^0\overline{K}^0 \to \pi^+\pi^-$ in this analysis.  This means that the amplitude factor 
$a_{K\overline{K}}$ includes both contributions from 
$K^+K^- \to \pi^+\pi^-$ and
$K^0\overline{K}^0 \to \pi^+\pi^-$. 

The amplitude for $\pi^0\pi^0 \to \pi^+\pi^-$ in Eq.~\ref{eqn:amp_pi0pi0_picpic}
has a different isospin factor at $T_0^2$ compared to
the amplitude for $\pi^+\pi^- \to \pi^+\pi^-$ in Eq.~\ref{eqn:amp_picpic_picpic} and
different masses for $\pi^0$ and $\pi^+$. 
In our fits we assume the equity $d_{\pi^0\pi^0} = d_{\pi\pi}$.
The constant $c_{\pi\pi}$ also accounts for the point-like term in
Eq.~\ref{eqn:amp_pipi_const}, and is involved in I=2 term, 
Eq.~\ref{eqn:amp_picpic_I2}, that makes it different from 
the $c_{\pi^0\pi^0}$.
For this reason we consider the $\pi^0\pi^0 \to \pi^+\pi^-$ sub-mode separately
from $\pi^+\pi^- \to \pi^+\pi^-$.



\end{document}